\documentclass[twocolumn]{emulateapj}
\usepackage{epstopdf}
\usepackage{graphicx}

\shorttitle{Validation of KOIs}
\shortauthors{PLAVCHAN ET AL.}

\begin{document}

\title{Investigation of Kepler Objects of Interest Stellar Parameters from Observed Transit Durations}

\author{Peter Plavchan\altaffilmark{1}, Christopher Bilinski\altaffilmark{2}, and Thayne Currie\altaffilmark{3}}

\altaffiltext{1}{NASA Exoplanet Science Institute, California Institute of Technology, M/C 100-22, 770 South Wilson Avenue, Pasadena, CA 91125; plavchan@ipac.caltech.edu}
\altaffiltext{2}{University of Arizona}
\altaffiltext{3}{U Toronto}

\begin{abstract}
The Kepler mission discovery of candidate transiting exoplanets (KOIs) enables a plethora of ensemble analysis of the architecture and properties of exoplanetary systems.  We compare the observed transit durations of KOIs to a synthetic distribution generated from the known eccentricities of radial velocity (RV) discovered exoplanets.  We find that the Kepler and RV distributions differ at a statistically significant level.  We identify three related systematic trends that are likely due to errors in stellar radii, which in turn affect the inferred exoplanet radii and the distribution thereof, and prevent a valid analysis of the underlying ensemble eccentricity distribution.   First, 15\% of KOIs have transit durations $>$20\% longer than the transit duration expected for an edge-on circular orbit, including 92 KOIs with transit durations $>$50\% longer, when only a handful of such systems are expected.   Second, the median transit duration is too long by up to $\sim$25\%.  Random errors of $<$50\% in the stellar radius are not adequate to account for these two trends.  We identify that incorrect estimates of stellar metallicity and extinction could account for these anomalies, rather than astrophysical effects such as eccentric exoplanets improbably transiting near apastron.  Third, we find that the median transit duration is correlated with stellar radius, when no such trend is expected.  All three effects are still present, although less pronounced, when considering only multiple transiting KOI systems which are thought to have a low false positive rate.  Improved stellar parameters for KOIs are necessary for the validity of future ensemble tests of exoplanetary systems found by Kepler.
\end{abstract}


\section{Introduction}

The Kepler mission is revolutionizing our understanding of exoplanets \citep[]{borucki}, including among its many highlights the discovery of three terrestrial exoplanets orbiting the M dwarf Kepler Object of Interest (KOI) 961 \citep[]{muirhead}, and a 2.4 $R_\oplus$ exoplanet in the habitable zone of its host star \citep[]{borucki4}.  A list of 312 KOIs was published in \citet[]{borucki2}, derived from Q0-1 Kepler time-series data.  This was soon followed by a second list of 1235 KOIs in \citet[]{borucki3} derived from Q1-Q5 Kepler data.   The first two KOI releases relied on stellar parameters from the Kepler Input Catalog \citep[KIC,][]{brown}. \citet[]{batalha3} (hereafter B13) announced 2321 candidate transiting exoplanet KOIs orbiting 1783 host stars from an improved pipeline analysis of the Q1-Q6 Kepler data, and provided updated stellar parameters.  A list of $\sim$2700 candidates from analysis of the Q1-Q8 time-series was recently released by the Kepler team, but does not include new estimates of stellar parameters \citep[]{burke13}.   This large ensemble of exoplanet candidates, including many multi-exoplanet systems -- `multis' for short -- beckons for the ensemble analysis of exoplanetary system architectures \citep[]{fabrycky,howard,morton,plavchan,figueira,youdin}.

The ensemble analysis of KOIs and the determination of the frequency of Earth-sized planets ($\eta_\oplus$) in habitable zone orbits relies on the accuracy of the estimated stellar host parameters such as mass, radius and temperature \citep[]{batalha,batalha2,pin,traub}.  Recent studies suggest that the KIC mischaracterizes some objects, in particular the surface gravity for stars with effective temperatures less than $\sim$4500 K.  For example, \citet[]{mann,ciardi} demonstrate that M ``dwarf'' KIC targets brighter than a Kepler magnitude of 14th are most likely giants.  Additionally, \citet[]{muirhead2} report that M dwarf KOI host stars have over-estimated radii.  \citet[]{dressing} confirm this trend and find that the typical M dwarf radius in the Kepler sample is over-estimated by $\sim$30\%.   B13 acknowledges these limitations of the new and old stellar parameters, given that the observational methods used to characterize Kepler host stars were optimized for FGK stars.   In contrast, \citet[]{everett} find that KOI stellar radii are instead under-estimated for 87\% of their 268 sample of faint (Kepler magnitude $>$14) FGK host stars, including factors of more than $\sim$35\% for one-quarter of their sample. 

In this paper we use the transit durations of KOIs for two interdependent purposes: one, as a probe of the inferred eccentricity distribution of the candidate exoplanets in comparison to exoplanets discovered with the radial velocity (RV) technique, and two, as a diagnostic of the accuracy of the estimated stellar parameters.  The analytic framework for these two investigations is already laid out in \citet[][]{ford} (hereafter F08) and \citet[]{burke}. A comprehensive analysis is already carried out with the \citet[]{borucki3} KOI list by \citet[]{moorhead}.  This work presents a follow-up to the analysis in \citet[]{moorhead} for the more recent KOI list in B13.  Additionally, \citet[]{dawson} carry out a thorough modeling when the impact parameter can reliably be estimated for Jovian-sized exoplanet KOIs with high S/N transits. 

In ${\S}$2, we present the samples of RV exoplanets and KOI candidate exoplanets used for this investigation.  In ${\S}$3, we present the parameter we calculate for the anomalous transit duration.  In ${\S}$4 we present how we generate a synthetic population of transiting exoplanets from the population of RV-discovered exoplanets.  In ${\S}$5, we present our results from a comparison of the synthetic population of transiting exoplanets with the observed KOIs and subsets thereof.  In ${\S}$6 we discuss the contribution of detection biases and false positives to our results, we assess the accuracy of KOI stellar parameters from our results, and we comment of the dependence of eccentricity of exoplanet radius and multiplicity.  In ${\S}$7 we present our conclusions.

\section {Samples}

\subsection{KOIs}

We use the second and third tabulation of KOI candidate exoplanet and stellar parameters from \citet[]{borucki3} and B13 respectively; the latter is available at the NASA Exoplanet Archive \citep[]{akeson}.   We make subsets of terrestrial, Neptune-like, and Jovian KOIs from B13 at exoplanet radii of $R_{pl}<$2, 2$<R_{pl}<$6, and $R_{pl}>$6 $R_\oplus$ earth radii respectively.  We also separately consider singles and multis: $\sim$10-35\% of single exoplanet KOIs may be false-positives \citep[]{morton,sophie,colon}, whereas the false positive rate for multis is thought to be very small \citep[]{lissauer}.  By comparing these latter two samples, we can assess the role of false positives in our analysis.  We treat multiple KOIs orbiting the same host star as independent statistical tests. 

The KOIs primarily orbit FGK host stars.  We do not filter KOIs by their host mass, temperature, or radius, since we are interested in identifying any systematic trends as a function of stellar parameters. We restrict ourselves to KOIs in B13 with transits detected with S/N$>$10 where previous work has shown the KOI list to be reasonably complete \citep[]{howard,Christiansen}.  We also restrict ourselves to KOIs in B13 with planet orbital periods $P<160$ days, corresponding to 3 or more observed transits.  Our final sample comprises 2205 of the 2321 KOIs in B13. We consider the impact of KOI detection biases on our analysis in ${\S}$4 \& 6.

\subsection{RV-Discovered Exoplanets}

Next, we use the period and eccentricity values of 164 published RV-discovered exoplanets as listed at the NASA Exoplanet Archive as of March 7th, 2012 \citep[]{akeson}.  We exclude RV exoplanets with periods greater than 160 days and M$_p sin$i$>$30 M$_{J}$.  The mass constraint for the definition of an ``exoplanet'' is the same loose criteria as adopted by the NASA Exoplanet Archive and other major exoplanet archives \citep[]{wright}.  Nearly 500 exoplanets have been discovered with the RV method, but the majority possess orbital periods greater than 160 days and/or no constraints on the eccentricity.  The mean eccentricity for this 164 exoplanet sample is 0.18.   Whereas the distribution of periastron angles for RV exoplanets is uniform random on a sphere, the distribution of periastron angles for transiting exoplanets will more strongly favor angles perpendicular to the plane of the sky for eccentric planets due to the increased probability of transit \citep[F08,][]{burke}.  Thus, we do not use the measured periastron angles for RV planets as in \citet[]{kane} since they have an incorrect distribution for comparison to the transiting planets.  

We assume that this RV sample represents a viable control population to compare to KOIs, without reverting to analytic or model planet formation synthesis eccentricity distributions. There are several limitations to this assumption.  The stellar mass distribution for RV discovered exoplanets is centrally peaked around one solar mass, which is qualitatively similar but not identical to the stellar mass distribution of KOIs; the RV distribution of stellar masses is relatively broader.  Also, only $\sim$4\% of KOIs have radii $>$1 $R_J$, and only 5\% of our RV-discovered sample has masses $<$1 $M_J$.  Thus, while we are comparing two sets of exoplanets, the two populations likely possess different bulk densities, composition, and dynamical histories that we do not account for.  Next, we ignore differences due to Galactic location -- Kepler systems are at larger distances from the Earth than RV-discovered exoplanets, and neither sample is magnitude nor volume-complete. We discuss additional detection biases inherent in this RV sample in ${\S}$4.   

\section{Observed Transit Duration Anomalies}

Equal to $\tau_0$ in F08 in the limit $R_{pl}/R_*\rightarrow 0$, we define the transit duration anomaly dimensionless parameter $\alpha$:

\begin{equation}
\alpha = \frac{a T \pi}{(R_* + R_p)P} = \left(\frac{\pi G M_*}{4 P}\right)^{1/3} \frac{T}{(R_* + R_p)}
\end{equation}

\noindent where $\alpha$ is the ratio of the observed transit duration to the expected transit duration for a circular edge-on orbit, and can be computed from the KOI table parameters in B13.  $R_*$ and $R_p$ are the stellar and exoplanet radii, $P$ is the orbital period, $T$ is the duration of the transit, $a$ is the semi-major axis of the exoplanet, $G$ is Newton's gravitational constant, and $M_*$ is the stellar mass.  The semi-major axis of the exoplanet $a$ is calculated from the stellar mass $M_*^{1/3}$ and $P$ via Kepler's 3$^{rd}$ law.    $P$ and $T$ are quantities measured from the time-series, whereas $R_*$ and $M_*$ (and consequently $a$) are estimated from ancillary observations and models such as observed colors, spectroscopy and theoretical isochrones \citep[]{batalha,batalha2,batalha3,muirhead2}.  Thus, the measurement of $\alpha$ depends strongly on the determinations of $T$ and $R_*$, and weakly on $P$, $M_*$, and $R_p$ (since $R_p\ll R_*$).  The error budget for $\alpha$ is in turn dominated by uncertainties in the stellar radius. This is because errors in the observed KOI transit durations $T$ are typically $<1\%$, the errors in the observed period $P$ are also generally known to better than 1 part in 10$^3$ (B13), $R_p\ll R_*$, and due to the weak dependence on stellar mass in the semi-major axis $a$.  Values for $\alpha$ are tabulated in Table 1 for all KOIs from B13.   

From Equation 1, we can see that the observed transit duration in hours subtracted from the expected transit duration in hours for a circular edge-on orbit has a dependence on orbital period -- transit durations are longer at longer orbital periods, all else being the same:

\begin{equation}
\Delta t \mbox{(hrs)} = T_{obs} - T_{circ} = (\alpha -1) \frac{P(R_*+R_p)}{\pi a}\propto \sqrt{a}
\end{equation}

\noindent The upper envelop of $\Delta t$ values in Figure 1 of \citet[]{kane} follows the expected $\sqrt{a}$ dependence as a consequence of Kepler's 3$^{rd}$ Law, and thus $\Delta t$ is a not a valid quantity to investigate eccentricity distributions over a range of orbital periods.  Our expression in Equation 1 is dimensionless with no unbalanced dependence on orbital period save for the intrinsic eccentricity distribution.  
  
For an eccentric, non-edge-on orbit, the square of the transit duration anomaly can be written in a similar fashion to the square of Equation 1 in F08 as:

\begin{equation}
\alpha^2 \approx \left(\frac{d_t}{a}\right)^2 \left(\frac{1}{1-e^2}\right)\left(1-\frac{b^2}{(1+R_p/R_*)^2}\right) 
\end{equation}

\noindent where $e$ is the eccentricity,  $b$ is the impact parameter, and $d_t$ is the exoplanet-star separation during transit as defined and derived in F08.  Assuming as in F08 that $a/R_*  \gg 1$, $a/R_p  \gg 1$, and $T/P  \ll 1$, $d_t$ can be assumed to be approximately constant during transit (hence the approximation in Equation 3).   

From Equation 3, we can identify potential eccentric exoplanets with values of $\alpha\ll1$ transiting near periastron or $\alpha\gg1$ transiting near apastron.  Since $b$ is degenerate with eccentricity for a given KOI, the KOIs with $\alpha\ll1$ may alternatively be grazing transits.  We tabulate these candidates in Table 2.  KOIs with $\alpha\gg1$ are low-probability occurrences (${\S}$5.1), and we exclude KOIs with KOIs with $\alpha\gg1$ from Table 2 unless $\log [Fe]/[H]>-0.11$ and $A_V < 0.33$ mag (${\S}$6.3).

The four KOIs with the smallest values of $\alpha$ -- KOIs 338.01, 338.02, 977.01, and 1054.01 -- are reported to orbit giant stars with log $g<2.5$ with periods of 1.4--7 days interior to the estimated stellar radii, and may instead be associated with photospheric activity or a blend.  Visual inspection of the time-series for KOIs 977 and 1054 support this hypothesis.  The KOI 338 system has two candidate exoplanets, and the light curve exhibits clear transit-like dips that are short w/r/t to the transit period, and is specifically mentioned in B13 and \citet[]{fabrycky}.   We exclude these four KOIs from Table 2.

\section{Generating A Synthetic Population of Eccentric Transiting Exoplanets}

Equations 1 and 3 do not permit a straightforward computation of a KOI eccentricity from its observed transit duration, since the impact parameter $b$ and periastron angle $\omega$ are generally unconstrained and degenerate with the eccentricity for a given KOI exoplanet.  High S/N Jovian KOIs are an exception since their light curves with resolved ingress and egress slopes and durations are amenable to model fitting to accurately constrain $b$ and $\omega$ as in \citet[][]{dawson}.  \citet[]{burke} and F08 demonstrate that in the absence of constraints on $b$ and $\omega$, the ensemble distribution of $\alpha$ values can instead be used as a proxy for the ensemble eccentricity distribution. Thus, we convert the known distribution of eccentricities of confirmed RV exoplanets into a simulated distribution of transiting exoplanet $\alpha$ values to compare to the KOI $\alpha$ distribution.  We assume that every RV exoplanet ``transits'' with a distribution of $b$ and $\omega$ values that are properly weighted by the transit probability.  The angle of periastron cannot be assumed to be uniform random on a sphere as is true for RV discovered exoplanets and as is assumed in \citet[]{kane}.  Accurate probability distributions for $b$, $e$ and $\omega$ for transiting planets are given in \citet[][Equations 14--16; Figure 4]{burke}, and we numerically integrate these probabilities over $b$ and $\omega$ using the prescription in \citet[]{burke} combined with the known RV eccentricities.

In principle, we can make a straightforward comparison of the simulated distribution of $\alpha$ values from RV-discovered exoplanets to the empirical distribution of $\alpha$ values for KOIs.  If they match, as assessed by a 1-D two sample Kolmogorov-Smirnov (K-S) test, then one can assert that the underlying eccentricity distributions are the same.  However, before we proceed we must first attempt to account for uncertainties and detection biases between the two methods.  First, the stellar parameter uncertainties dominate the error budget for $\alpha$ in Equation 1 (${\S}$3).    To assess the impact of stellar parameter uncertainties on KOI $\alpha$ values, we include 20\% and 50\% Gaussian random errors in our RV simulated distribution of $\alpha$ values, which could represent realistic 20 or 50\% errors in the stellar radius, or alternatively $>$70\% errors in the stellar mass given the weaker one-third dependence of $\alpha$ on stellar mass. 

Second, we correct for the difference in detection sensitivity (survey completeness) as a function of orbital period for the transit and RV methods, because the eccentricity distribution of RV-discovered exoplanets has a relatively strong dependence on orbital period due to tidal circularization \citep[]{butler,rasio}.  For the RV method, the detection probability falls off as the semi-major axis $a^{-1/2}$, whereas for the transit method the detection probability falls more rapidly as $a^{-1}$.  For example, 80\% of our sample of RV-discovered exoplanets have orbital periods $P<$ 73 days, whereas 80\% of KOIs have $P<$ 29 days, a difference of more than a factor of two.  We plot the cumulative distribution functions (CDFs) as a function of orbital period $P$ for the RV-discovered exoplanets, KOIs and various subsets thereof in Figure 1.  

Instead of generating standard CDFs as a function of $\alpha$, where each increment in the CDF value is 1/N and N is the sample size (N=164 from ${\S}$2.2), we generate weighted CDFs where each increment is $w_i$ and $\sum_{i=1}^N w_i = 1$. We calculate the $w_i$ values to yield a weighted CDF($P$) for the simulated exoplanets that is identical to the standard unweighted CDF($P$) for the KOIs or sub-sets thereof.  In practice, this means that shorter orbital period simulated exoplanets and their associated $\alpha$ values are typically given more weight than the longer period simulated exoplanets.  The particular weights for a given simulation depend on the particular period distribution of the KOIs and subsets thereof to which the simulation is being compared.  For example, KOIs with $R_{pl}<$2 $R_\oplus$ generally have shorter orbital periods than KOIs with $R_{pl}>$6 $R_\oplus$ due to the relative detection incompleteness for the smaller radius KOIs at longer orbital periods.  By weighting the $\alpha$ values in this manner, we account for this relative incompleteness as well.  Our approach is equivalent to generating a standard CDF from a Monte Carlo simulation where planets are drawn non-randomly from the RV-discovered set to yield the desired CDF($P$), while fixing the number of simulated planets at $N$ to retain the applicability of the K-S test in ${\S}$5.  The net result is a simulated and weighted CDF($\alpha$) to compare to the unweighted CDF($\alpha$) for KOIs or a subset thereof, where the underlying period distributions are identical between the two samples. 

Third, the RV technique is also biased against detecting highly eccentric planets for a sparse cadence of observations.  Correcting for this bias would increase the simulated mean eccentricity. Since the correction would strengthen the results presented in ${\S}$5, it can thus be safely ignored.   We also do not correct for the $sin$i degeneracy in exoplanet mass, nor the bias of both transit and RV techniques to preferentially detect higher mass/radius planets, nor the lack of sensitivity of the RV technique to detect terrestrial planets.  The dependence of eccentricity on exoplanet mass is not observationally well constrained for RV-discovered planets without reverting to theoretical formation models, and thus it is difficult to correct for a priori.   

Finally, we do not correct for the qualitatively similar but distinct distributions as a function of stellar mass between the KOIs and RV-discovered exoplanets; there is no published observational literature presenting a dependence of exoplanet eccentricity on stellar mass, and no trend is apparent to us in our data from the NASA Exoplanet Archive \citep[]{akeson}.  To summarize, we assume that the sample of eccentricities of RV-discovered exoplanets serves as an appropriate proxy for a `control sample' to compare to the KOIs, and in particular the KOIs with $R_{pl}>$6 $R_\oplus$, after simulating $\alpha$ values from these eccentricities, accounting for differences in the period distributions, and accounting for stellar parameter uncertainties.  We defer the discussion of additional detection biases for the KOIs until ${\S}$6.1.

\section{Results}

In this section, we first present the CDFs of simulated $\alpha$ values from the RV-discovered exoplanets in ${\S}$5.1, which are then used as a benchmark to compare to the empirical CDFs for KOIs in ${\S}$5.2.

\subsection{Simulated Distributions}

In Figure 2, we plot the CDFs of simulated $\alpha$ values from the RV-discovered exoplanet sample. We tabulate the distribution medians $\tilde{\alpha}$ and two-sided standard deviations $\pm\sigma$ in Table 3.  Our simulated $\alpha$ distribution -- with an average eccentricity of 0.18, no added random errors and no period distribution correcting weights -- has a median, mean and standard deviation in $\alpha$ of $(\tilde{\alpha},\mu,\sigma)=(0.73,0.69,0.25)$.  For comparison, the model in F08 that best matches known RV-discovered exoplanets is a Rayleigh eccentricity distribution with an average eccentricity of $\sim$0.25 and with a Rayleigh parameter R(0.3) \citep[]{juric}.  This F08 model produces a distribution in $\alpha$ with $(\mu,\sigma)\sim(0.74,0.29)$ that is comparable to the implementation of our simulations.

As previously mentioned in ${\S}$3, a large value of $\alpha\gg1$ (assuming accurate stellar parameters) requires a high eccentricity and transit near apastron, which is a low probability occurrence.  This explains the lack of simulated $\alpha$ values $>$1.2 when no additional random errors are included \citep[see also][Figure 4]{burke}.  For the addition of 20\% random errors in measuring $\alpha$, and when the simulated period distribution is weighted to match that of all KOIs in B13, we find 2.4\% of $\alpha$ values $>$1.2 and 0\% with $\alpha>$1.5.  For 50\% errors, we find 10.4\% of $\alpha$ values $>$1.2, and 2.4\% with $\alpha>$1.5.   Thus, our simulations show that significant random errors $>$50\% in measuring $\alpha$ can produce a false overabundance ($\sim$10\%) of eccentric planets improbably transiting near apastron.  However, the median $\tilde{\alpha}$ values of the synthetic distributions are relatively insensitive to the amplitude of the random errors (Table 3).  We next compare the simulated distributions to the empirical KOI $\alpha$ distributions.

\subsection{Empirical Distributions}

In Figure 2 we also plot the CDFs of $\alpha$ values for all KOIs from B13 and for KOIs with $R_{pl}<2 R_\oplus$, $2 R_\oplus< R_{pl} <6 R_\oplus$, and $R_{pl}>6 R_\oplus$, for single KOI systems, and multiple KOI systems (`multis') also all from B13.  The $\alpha$ distribution for all KOIs is significantly different from the corresponding RV $\alpha$ distribution in ${\S}$5.1, in disagreement with \citet[]{kane}.   This is particularly evident when comparing the distribution medians, which can deviate by up to 25\% between the empirical and simulated distributions.  This result suggests that there is a systematic under-(over-)estimate of stellar radii (mass), since the transit durations and periods are known to better than 1\% and 0.1\% respectively (Equation 1, ${\S}$3).  In particular, the CDFs for KOIs with $R_{pl}<6 R_\oplus$ appear to be unphysical.  A population of exoplanets on circular orbits only, and with uniform random impact parameters, produces an $\alpha$ distribution with $\tilde{\alpha}=0.866$.  As the average eccentricity is increased from zero, $\tilde{\alpha}$ decreases.  Thus, any $\tilde{\alpha}>0.866$ is unphysical, implying the KOIs typically have ``more circular than circular'' orbits with $\tilde{\alpha}>0.9$.  To ascertain the validity of these claims, we discuss in ${\S}$6 the relevance of detection biases and false positives in the list of KOIs.

Formally, we evaluate the K-S test statistic between all distributions, and the K-S test probabilities are listed in Table 4.  The simulated $\alpha$ distribution from RV-discovered exoplanets, with no additional random errors, has a probability of being drawn from the same parent population as the KOIs in B13 with $R_{pl}>6 R_\oplus$ ($2 R_\oplus<R_{pl}<6 R_\oplus$,$R_{pl}<2 R_\oplus$, all) of $4.48\times10^{-4}$ ($3.91\times10^{-12}$,$3.90\times10^{-23}$,$3.31\times10^{-16}$).   Thus, KOIs with $R_{pl}>6 R_\oplus$ most closely resemble the simulated $\alpha$ distribution, but are still statistically distinct at the $>3\:\sigma$ level.   For KOIs with $R_{pl}<6 R_\oplus$, these results hold at high statistical significance for additional errors in the stellar radii.  For KOIs with $R_{pl}>6 R_\oplus$, the K-S test probability increases to $8.63\times10^{-4}$ for 20\% errors and to $\sim$2.4\% for 50\% errors.  From Figure 2, it is apparent that much of the disagreement for KOIs with $R_{pl}>6 R_\oplus$ is for large values of $\alpha$.  However, the disagreement in CDFs is noticeable at all values of $\alpha$ for $R_{pl}<6 R_\oplus$.

Next, there is a higher occurrence rate for all KOIs with $\alpha>$1.5 than expected for both a Rayleigh eccentricity distribution (F08) and for our simulated distributions with no added random errors.  This is true for all KOI radii, whether in a single or multiple system, although the percentages are smaller for the multis.   For all KOIs, we find 15\% have $\alpha>$1.2, and 4\% with $\alpha>$1.5.  These observed percentages are larger than but comparable to the $\alpha$ distribution simulated from RV-discovered exoplanets with 50\% Gaussian random errors in measuring $\alpha$ as presented in ${\S}$5.1.  For the single KOI systems, the percentages are 16.3\% and 5.7\% for $\alpha>$1.2 and 1.5 respectively.  For multis, the corresponding percentages are smaller at 11.6\% and 1.6\% respectively, a difference also noted in \citet[]{moorhead}.

This excess of KOIs with $\alpha>$1.5 is also noted in \citet[]{moorhead} and \citet[]{wang} for the \citet[]{borucki3} tabulation of KOIs.  We confirm that this excess is still present for KOIs in B13.  If we assume that our RV and KOI samples are drawn from the same population of exoplanets, these KOIs must be possess significant systematic errors in the stellar parameters of $\sim$20--50\%.   In ${\S}$6.3, we identify how the stellar radius (mass) could be under-(over-)estimated by factors exceeding 1.5 (1.5$^3$) to correct some of these $\alpha$-values.  A blanket rejection of these KOIs is not recommended since that could inadvertently exclude a rare and genuine highly eccentric exoplanet transiting near apastron.   

\section{Discussion}

\subsection{Detection Biases}

In ${\S}$4, we correct for differences in the orbital period completeness of KOIs and subsets thereof through the use of weighted CDFs. The effect of this correction is minimal as shown in Figure 2 and Tables 3 and 4.   We now discuss two additional detection biases that are specific to KOIs, and we evaluate the impact of these biases on our results in ${\S}$5.  We expect a deficiency of transiting KOIs at high impact parameters due to two effects -- the lower S/N for shorter transit durations, and the manual removal of `V'-shaped grazing transits because of the increased probability of false-positives from stellar eclipsing binaries \citep[B13,][]{blender,Christiansen}.  It is beyond the scope of this work to obtain an accurate measure of the incompleteness due to these two biases.  Such an effort requires a detailed modeling of the Kepler pipeline recovery of injected synthetic transits, which fortunately is undertaken in \citet[]{Christiansen}.  \citet[]{Christiansen} find there is no significant bias in the Kepler pipeline recovery of individual transit events (known as Kepler Threshold Crossing Events, or TCEs) as a function of transit duration.  However, the \citet[]{Christiansen} result does not account for any human biases introduced in the promotion to a KOI from the Kepler TCEs. 

We instead use the impact parameter $b_{circ}$ from B13 to provide a reasonable estimate of incompleteness at short transit durations.   $b_{circ}$ is calculated assuming a circular orbit and is determined from the ingress and egress times after accounting for limb-darkening \citep[]{seager}. We plot the distribution of KOIs a function of $b_{circ}$ in Figure 3.  The frequency of KOIs with $b_{circ}>0.9$ is $\sim$45\% smaller than KOIs with $0.8<b_{circ}<0.9$, particularly for KOIs with $R_{pl}<6 R_\oplus$.  No such decrease would be expected for a uniform random impact parameter for circular orbits, and the number of KOIs with $b_{circ}>0.9$ should in fact slightly increase for a population of eccentric orbits.  Thus, there is a noticeable and measurable incompleteness of KOIs with $b_{circ}>0.9$.  For reference, a transit duration for a circular orbit at an impact parameter of $b$=(0.5,0.6, 0.7, 0.8, 0.9) is (0.866,0.8,0.71,0.6, 0.44) times as long as an edge on ($b$=0) transit.

For an approximate lower bound, we estimate that 50\% (240) of KOIs are missing with $b_{circ}>0.9$, primarily with $R_{pl}<6 R_\oplus$, by linearly extrapolating from the number of KOIs with $0.7<b_{circ}<0.8$ and $0.8<b_{circ}<0.9$. In other words, we assume for the lower bound that the list of KOIs is complete for $b_{circ}<0.9$ and the trend in $b_{circ}$ frequency is linear from 0.7--1.0.  Similarly, for an approximate upper bound, we estimate that 40\% (471) of KOIs are missing with $b_{circ}>0.8$ by linearly extrapolating from the number of KOIs with $0.6<b_{circ}<0.7$ and $0.7<b_{circ}<0.8$. In other words,  we assume for the upper bound that the list of KOIs is complete for $b_{circ}<0.8$ only, and the trend in $b_{circ}$ frequency is linear from 0.6--1.0).  This fraction of missing high impact parameter KOIs corresponds to $\sim$10--20\% of all KOIs in B13.   The missing KOIs would have values for $\alpha$ less than $\sim$0.3 for KOIs with $R_{pl}>6 R_\oplus$ and less than $\sim$0.1 for KOIs with $R_{pl}<2 R_\oplus$ (Equations 1,3).    Adding this population of missing KOIs to our sample would effectively compress the existing KOI CDFs plotted in Figure 2 from the vertical range of (0,1) to $\sim$(0.09,1) or (0.167,1) for 10 and 20\% incompleteness respectively.  Thus, we can estimate a corrected median value of $\tilde{\alpha}_c$=0.876--0.906 for all KOIs.  This scenario remains inconsistent at a statistically significant level with a population of RV exoplanets with a mean eccentricity $\bar{e}=$0.18 and $\tilde{\alpha}$=0.73 from ${\S}$4.  It also remains marginally inconsistent in an unphysical fashion with a population of exoplanets on only circular orbits ($\bar{\alpha}=0.866$).   Further, the distribution of $b_{circ}$ values in Figure 3 implies an average KOI eccentricity of $e>$0 and thus $\tilde{\alpha}_c<0.866$, since $\bar{b}_{circ}>=0.7$ even before correcting for the incompleteness at $b_{circ}>0.8$, and since an average value of $\bar{b}_{circ}=$0.5 would be expected for a population of exoplanets with only circular orbits.   Thus, the estimated incompleteness of KOIs at short transit durations is unable to explain the differences between the simulated and empirical $\alpha$ distributions, in particular accounting for the KOI median values of $\tilde{\alpha}$.  

Tackling the question of incompleteness at short transit durations from a different direction, we can ask -- if the median $\tilde{\alpha}$ for KOIs should be equal to the median $\tilde{\alpha}$ for the simulated exoplanets such that they are drawn from the same parent population, what fraction of KOIs with $b_{circ}>0.9$ (0.8) would be missing to account for the observed discrepancies in $\tilde{\alpha}$?  For all KOIs, the percentage would be 58.4\% (1288 KOIs), corresponding to $\sim$4.7 (1.9) times the existing number of KOIs with $b_{circ}>0.9$ (0.8), or approximately $\sim$5 (2.7) times our estimated incompleteness in the preceding paragraph.  Broken down by KOI subsets, for multis, singles, KOIs with $R_{pl}>6 R_\oplus$, $2 R_\oplus<R_{pl}<6 R_\oplus$, and $R_{pl}<2 R_\oplus$, the corresponding percentages are 65,52,17,54, and 75\% times the total number of KOIs, corresponding to factors of 5.2, 4.1, 1.4, 4.4, and 6 times the existing number of KOIs with $b_{circ}>0.9$ respectively.   These incompleteness factors would appear to be inconsistent with the observed distribution for $b_{circ}<$0.9 in Figure 3 for all KOI radii.  We can conclude that there are instead likely biased errors present in the stellar parameters.

\subsection{False Positives}

False-positives in the B13 KOI list are reported at the $\sim$10-35\% level  \citep[]{morton,sophie}.   \citet[]{colon} finds no significant correlation in the false positive rate with exoplanet radius and stellar effective temperature, albeit from a limited sample.    Fortunately, multiple exoplanet KOI systems are thought to have a very low false-positive rate of a few percent or less \citep[]{lissauer}.   Thus, we can compare single exoplanet KOIs to multis to assess the impact of false-positives on the transit duration anomalies $\alpha$.   As can be seen in Figure 2 and Tables 3 and 4, the difference in the median $\tilde{\alpha}$ for singles and multis is marginal -- 0.93 vs 0.94 respectively -- both before and after correcting for differences in the orbital period distributions.  Thus the false positive rate is unable to account for the errantly high values of $\tilde{\alpha}$ for KOIs when compared to the simulated distributions.  

However, we do see a marked difference between singles and multis in the frequency of KOIs with $\alpha>$1.2 and 1.5 as noted in ${\S}$5.2.   Taking the ratio of  percentages, false positives can account for $\sim$30\% of KOIs with $\alpha>$1.2, and $\sim$70\% of all KOIs with  $\alpha>$1.5.  Thus a fraction of KOIs with abnormally large $\alpha$ values can be accounted for by false positives, but not all.  The remaining $\sim$12\% and $\sim$2\% of all KOIs with $\alpha>$1.2 and 1.5 respectively must still be accounted for.  The statuses of the KOIs with $\alpha>$1.2, particularly for single exoplanet KOI systems, need to be confirmed in future work to enable a scientifically valid comparison of the eccentricity distributions of RV and Kepler planets.   

\subsection{Accuracy of Stellar Parameters Inferred from Transit Duration}

The significant differences between the $\alpha$ distributions for RV exoplanets and Kepler candidates raises the question about the validity of the KOI host stellar parameters, which we now turn to discuss.  From Figure 2 and ${\S}$5, we find that Gaussian random errors in measuring $\alpha$ of $\sim$20--50\% can explain the overabundance of KOIs with $\alpha>$1.2, after accounting for false positives in ${\S}$6.2.   However, 20--50\% Gaussian random errors tend to over-predict the frequency of KOIs with $\alpha<$0.4.  Additionally, the median $\tilde{\alpha}$ values of KOIs and subsets thereof are up to 25\% larger than their simulated counterparts, with the exception of Jovian KOIs with $R_{pl}>6 R_\oplus$.  Our simulations show that the median $\tilde{\alpha}$ value is roughly independent of the simulated Gaussian random uncertainties.  Our analysis in ${\S}$6.1 also shows that neither the correction for the orbital period distributions, nor the estimated correction for the incompleteness at high impact parameters, can fully explain the discrepancies in the median values of $\tilde{\alpha}$.       Thus, rather than Gaussian random errors, systematic (meaning one-sided, non-Gaussian) over-estimates in the measurement of $\alpha$ of up to $\sim$25\% on average, and up to 50\% for individual KOIs, are likely required to reconcile the observed KOI $\alpha$ distributions with their simulated counterparts.


Given the relatively small uncertainties in the KOI orbital periods and transit durations of $<$0.1\% and $<$1\% respectively in B13, the errors in measuring $\alpha$ must either be due to errors in the stellar radii or the one-third power of the stellar masses.  A $>$20\% under-estimate in the cube root of the stellar mass corresponds to a $>$70\% error in the stellar mass.  This would seem to be less plausible than an equivalent 20\% error in stellar radius, since stellar mass errors of $>70\%$ would require correspondingly large spectral type errors.  Thus, we conclude that systematic under-estimates of $\sim$20--50\% in the estimated stellar radii are the most likely explanation for the systematic over-estimates in the measurement of $\alpha$.   This conclusion is independently confirmed with spectroscopy of faint KOI host stars in \citet[]{everett}, who also find that the KOI stellar radii are systematically under-estimated by up to $\sim$30\%.  Thus, some KOIs identified as main sequence stars are likely to be more distant sub-giants.

Can we find additional indicators that there are systematic errors in the stellar radii that can explain our results?  In Figures 3--5 we plot the dimensionless $\alpha$ parameter as a function of stellar mass, radius and effective temperature, for both the \citet[]{borucki2} and B13 lists of KOIs, and for singles and multis from B13.  We bin the $\alpha$ values for each stellar parameter, with bins of 250 K, 0.1 $M_\odot$, and 0.1 $R_\odot$.  Our results are insensitive to adjustments in the bin width.  We derive median $\tilde{\alpha}$ values and quartile ranges for each bin, which are over-plotted in Figures 3--5 and show distinct trends as a function of stellar mass and radius, but not stellar effective temperature.  To quantify these trends, we perform a linear regression fit to the median values, excluding bins with fewer than 4 KOIs.  The linear coefficients and the standard errors from the fits are listed in Table 5. We identify a statistically significant trend for increasing stellar radius (mass) of a -0.27 (-0.28) change in $\tilde{\alpha}$ per $R_\odot$ ($M_\odot$) at the $\sim$5-$\sigma$ (3-$\sigma$) level for both the B13 and \citet[]{borucki2} KOIs, as well as for both single and multi KOIs from B13 albeit at a lower statistical significance. 


The trends in $\alpha$ as a function of stellar mass and radius imply one of two scenarios. First, there could be a systematic error in these two parameters for the ensemble of KOIs. Relative to the median $\tilde{\alpha}$ value for 1.0 $R_\odot$ ($M_\odot$) KOIs, the ensemble radii (or masses) would be under- (over-)estimated by an additional $\sim$15\% ($\sim$50\%) at 0.5 $R_\odot$ ($M_\odot$), and the ensemble radii (or masses) would be over- (under-)estimated by $\sim$15\% ($\sim$50\%) at 1.5 $R_\odot$ ($M_\odot$). As previously discussed, it is less plausible that the ensemble stellar masses in B13 and the KIC are in error by $\sim$50\% at the low- and high- mass ends, leaving systematic ensemble errors in the stellar radii as the more plausible explanation for the $\tilde{\alpha}$ trends.  Additionally, the trend with stellar radius has a higher formal statistical significance.

The first scenario relies on the assumption that the eccentricity distribution of exoplanets is intrinsically independent of the stellar spectral type.  Alternatively, the systematic trend in $\alpha$ could instead be due to a real change in eccentricity distributions as a function of spectral type.  A $\sim$15\% change in $\tilde{\alpha}$ between 0.5 and 1.0 $M_\odot$ (or 1.0 and 1.5 $M_\odot$) would correspond to a $>$0.2 change in the average ensemble eccentricity.  For example, the trend in $\tilde{\alpha}$ could imply that the average eccentricity is $\tilde{e}\sim$(0,0.25, 0.6) for (0.5,1,1.5) $M_\odot$ host stars respectively as inferred from Figure 9 in F08.  However, such a large eccentricity dependence on stellar mass is not reported in the literature for the eccentricities of RV-discovered exoplanets.  We again conclude that the trends in the ensemble $\tilde{\alpha}$ values for KOIs as a function of stellar mass and radius are likely primarily due to systematic errors in stellar radii.  This result contradicts recent work done for M dwarf KOIs by \citet[]{dressing}, and we are suggesting instead that M dwarf radii for KOI host stars are typically under-estimated rather than over-estimated.  The spectroscopic work of \citet[]{everett} is consistent with our result.

Finally, the over-abundance of KOIs with $\alpha>$1.5 identified in ${\S}$5 is independent of stellar mass, radius and temperature, and is only partially explained by false-positives.  We compare KOI $\alpha$ values against every property calculated in the KOI tables in B13.  We find that KOIs with $\alpha>1.2$ are preferentially found around low metallicity KOIs with $\log [Fe]/[H]<$-0.11 for all planet radii, and for $A_V > 0.33$ mag for $R_{pl}>6 R_\oplus$ as shown in Figure 7.  In other words, there is a lack of KOIs with $\alpha>$1.2 at high metallicities and low $A_V$. While the extinction may be a crude proxy for brightness, there is no a priori reason to expect either of these two trends.  Thus we conclude that the over-abundance of KOIs with $\alpha>1.2$ is likely due to errors in the calculated stellar metallicities and possibly extinction for these sources, rather than constituting a genuine class of exoplanets with high eccentricities (improbably) transiting far outside of periastron.  These errors in metallicity and possibly extinction in turn can readily produce the under-estimates of the stellar radii for these KOIs.  These stellar hosts are likely more distant (sub-)giants with larger stellar radii and larger secondary companions, or alternatively are false-positives.  This is consistent with the result in \citet[]{colon} that both false-positives they identify are the faintest targets in their sample.

\subsection{Eccentricity as a Function of Planet Radius and Multiplicity}

KOIs with $R_{pl}>6 R_\oplus$ have a statistically significant smaller median $\alpha$ compared to KOIs with $R_{pl}<6 R_\oplus$ (Table 3, Figure 2).  The difference is less distinguishable but still present when comparing KOIs with $2 R_\oplus<R_{pl}<6 R_\oplus$ and $R_{pl}<2 R_\oplus$.  Similarly, KOIs in multiple systems also have a slightly larger median $\alpha$ value compared to single KOIs.  All of these differences become slightly more pronounced after we correct the period distributions of each of these subsets to all match the period distribution of KOIs via a weighted CDF($\alpha$) (${\S}$4), as we show in Figure 8.  These results would appear to imply that multis are on more circular orbits than single exoplanets, and that larger exoplanets typically have larger eccentricities.  The circularity of the multiple planet systems has been previously reported in \citet[]{fang} and \citet[]{fabrycky} due to the small mutual inclinations of KOIs.  Our results are consistent with their more robust conclusions.  We must be cautious in interpreting the eccentricity distribution as a function of planet radius as noted in ${\S}$6.1 and 6.2.  Stellar parameter systematic errors are still prevalent in the KOI $\alpha$ values, and there is some incompleteness at high impact parameters for the smallest radius KOIs.  However, even after accounting for these factors, it will be difficult to reconcile the median $\tilde{\alpha}$ values for KOIs with  $R_{pl}>6 R_\oplus$ and  $R_{pl}<6 R_\oplus$.  It is likely that our result points to a different dynamical origin for exoplanets with $R_{pl}>6 R_\oplus$ and $R_{pl}<6 R_\oplus$, but this needs confirmation from future work with improved stellar parameters, completeness and false positive rejection.

\section{Conclusions}

We have carried out an updated analysis of the transit duration anomalies with the list of Kepler exoplanet candidates in B13.  In particular, we looked at the KOI distribution of transit durations compared to what would be expected from the eccentricity distribution of RV-discovered exoplanets as a function of stellar host parameters.  We find three related systematic errors in the KOI stellar parameters that preclude a scientifically valid ensemble comparison of the two samples at this time.  The systematic biases in stellar parameters impact the inferred distributions of exoplanet properties, including radius and habitability.  Thus any determinations of $\eta_\oplus$, the frequency of Earth-sized planets in the habitable zone, etc., must be treated with caution.  

First, there is an over-abundance of KOIs with transit durations $>$20\% and $>$50\% longer than expected, implying that most of these KOIs most likely have significantly under-estimated stellar radii.  This confirms the result in \citet[]{everett}.  We identify that biases in the estimated metallicity and extinction may explain these systems.  

Second, we identify that the median transit duration for all spectral types is up to $\sim$25\% too long, a result that is not explainable by sample incompleteness at short transit durations that we estimate to be $\sim$10-20\% of all KOIs.  This is again most likely due to the systematic under-estimates of KOI stellar radii.

Third, we identify statistically significant trends in the average transit duration as a function of stellar mass and radius, which again are likely due to errors in stellar radii as a function of spectral type rather than an underlying trend in eccentricity distributions.  
 
We thank the anonymous referee for their constructive input on improving the clarity, presentation and content of this paper.  This research has made use of the NASA Exoplanet Archive, which is operated by the California Institute of Technology, under contract with the National Aeronautics and Space Administration under the Exoplanet Exploration Program.


\begin{deluxetable}{rlll}
\tabletypesize{\scriptsize}
\tablewidth{0pt}
\tablecolumns{4}
\tablecaption{KOI Transit Duration Anomaly and Eccentricity Estimates\tablenotemark{a}}
\tablehead{
\colhead{KOI} & \colhead{$\alpha$} & \colhead{$e_{p}$\tablenotemark{b}} & \colhead{$e_{a}$\tablenotemark{c}}  }
\startdata
1.01	&	0.589	&	0.485	&	\nodata	\\
2.01	&	0.666	&	0.386	&	\nodata	\\
3.01	&	0.903	&	0.102	&	\nodata	\\
4.01	&	0.444	&	0.671	&	\nodata	\\
5.01	&	0.467	&	0.642	&	\nodata	\\
5.02	&	0.776	&	0.248	&	\nodata	\\
7.01	&	1.22	&	\nodata	&	0.198	\\
10.01	&	0.710	&	0.330	&	\nodata	\\
12.01	&	1.12	&	\nodata	&	0.117	\\
13.01	&	0.658	&	0.396	&	\nodata	\\
\enddata
\tablenotetext{a}{Table 1 is published in its entirety in the electronic edition. A portion is shown here for guidance regarding its form and content.}
\tablenotetext{b}{For a transit assumed to occur at periastron with $b=0$.}
\tablenotetext{c}{For a transit assumed to occur at apastron with $b=0$.}
\end{deluxetable}

\begin{deluxetable}{rrrr}
\tabletypesize{\scriptsize}
\tablewidth{0pt}
\tablecolumns{4}
\tablecaption{Candidate Eccentric KOIs}
\tablehead{
\colhead{KOI} & \colhead{$\alpha$} & \colhead{$e_{p}$\tablenotemark{a}} & \colhead{$e_{a}$\tablenotemark{b}}  }
\startdata
1845.02	&	0.155	&	0.953	&	\nodata	\\
371.01	&	0.156	&	0.952	&	\nodata	\\
2519.01	&	0.178	&	0.939	&	\nodata	\\
2522.01	&	0.190	&	0.931	&	\nodata	\\
2287.01	&	0.199	&	0.924	&	\nodata	\\
403.01	&	0.206	&	0.918	&	\nodata	\\
1815.01	&	0.221	&	0.907	&	\nodata	\\
1164.01	&	3.90	&	\nodata	&	0.877	\\
2046.01	&	2.47	&	\nodata	&	0.719	\\
783.01	&	2.41	&	\nodata	&	0.7077	\\
\enddata
\tablenotetext{a}{For a transit assumed to occur at periastron with $b=0$.}
\tablenotetext{b}{For a transit assumed to occur at apastron with $b=0$.}
\end{deluxetable}

\begin{deluxetable}{llllll}
\tabletypesize{\scriptsize}
\tablewidth{0pt}
\tablecolumns{6}
\tablecaption{$\alpha$ Distribution Medians and Standard Deviations}
\tablehead{
\colhead{Sample} & \colhead{Additional} & \colhead{Period Distribution} &  \colhead{Median} & \colhead{+1 Standard Deviation} & \colhead{-1 Standard Deviation}\\
\colhead{Distribution} & \colhead{Random Error} & \colhead{Corrected To} &  \colhead{ $\tilde{\alpha}$} & \colhead{ $+\sigma$} & \colhead{ $-\sigma$} 
}
\startdata
\multicolumn{6}{c}{KOIs} \\
\hline
All & 0\% & no correction  & 0.935   &   0.251  &    0.304 \\
$R_{pl} < 2 R_\oplus$ & 0\% & no correction & 0.986 & 0.245 & 0.244 \\
$2 R_{\oplus}<R_{pl}<6 R_{\oplus}$ & 0\% & no correction & 0.919 & 0.245 & 0.305 \\
$R_{pl}>6 R_{\oplus}$ & 0\% & no correction & 0.783 & 0.355 & 0.344 \\
singles & 0\% & no correction & 0.926 &  0.284 & 0.331 \\
multis & 0\% & no correction & 0.942 & 0.205 & 0.266 \\
$R_{pl} < 2 R_\oplus$ & 0\% & all KOIs & 0.999 & 0.238 & 0.246 \\
$2 R_{\oplus}<R_{pl}<6 R_{\oplus}$ & 0\% & all KOIs & 0.903 & 0.242 & 0.320 \\
$R_{pl}>6 R_{\oplus}$ & 0\% & all KOIs & 0.746 & 0.398 & 0.317 \\
singles & 0\% & all KOIs & 0.928 & 0.280 & 0.325 \\
multis & 0\% & all KOIs & 0.942 & 0.198 & 0.270 \\
\hline
\multicolumn{6}{c}{RV simulations} \\
\hline
All & 0\% & no correction & 0.730 & 0.217 & 0.286 \\
All & 0\% & All KOIs & 0.747 & 0.200 & 0.234 \\
All & 0\% & $R_{pl} < 2 R_\oplus$ KOIs & 0.766 & 0.182 & 0.253 \\
All & 0\% & $2 R_{\oplus}<R_{pl}<6 R_{\oplus}$ KOIs & 0.747 & 0.198 & 0.234 \\
All & 0\% & $R_{pl}>6 R_{\oplus}$ KOIs & 0.721 & 0.227 & 0.233 \\
All & 0\% & single KOIs & 0.747 & 0.201 & 0.240 \\
All & 0\% & multi KOIs & 0.766 & 0.179 & 0.253 \\
All & 20\% & All KOIs & 0.720 & 0.215 & 0.196 \\
All & 20\% & $R_{pl} < 2 R_\oplus$ KOIs & 0.709 & 0.270 & 0.308 \\
All & 20\% & $2 R_{\oplus}<R_{pl}<6 R_{\oplus}$ KOIs & 0.746 & 0.255 & 0.268 \\
All & 20\% & $R_{pl}>6 R_{\oplus}$ KOIs & 0.741 & 0.198 & 0.337 \\
All & 20\% & single KOIs & 0.670 & 0.349 & 0.194 \\
All & 20\% & multi KOIs & 0.704 & 0.271 & 0.260 \\
All & 50\% & All KOIs & 0.690 & 0.436 & 0.474 \\
All & 50\% & $R_{pl} < 2 R_\oplus$ KOIs & 0.716 & 0.364 & 0.383 \\
All & 50\% & $2 R_{\oplus}<R_{pl}<6 R_{\oplus}$ KOIs & 0.649 & 0.590 & 0.366 \\
All & 50\% & $R_{pl}>6 R_{\oplus}$ KOIs & 0.747 & 0.615 & 0.483 \\
All & 50\% & single KOIs & 0.603 & 0.498 & 0.324 \\
All & 50\% & multi KOIs & 0.799 & 0.399 & 0.506 \\
\enddata
\end{deluxetable}

\begin{deluxetable}{llllll}
\tabletypesize{\scriptsize}
\tablewidth{0pt}
\tablecolumns{6}
\tablecaption{K-S test probabilities}
\tablehead{
\colhead{Sample One} & \colhead{Sample One} &  \colhead{Sample Two} 	& \colhead{Additional} & \colhead{Sample Two} & \colhead{Probability} \\
\colhead{} & \colhead{Period Distribution} &  \colhead{} 		& \colhead{Random Error} & \colhead{Period Distribution} & \colhead{}\\
\colhead{} & \colhead{Corrected To} &  \colhead{} 			& \colhead{Sample Two\tablenotemark{a}} & \colhead{Corrected To} & \colhead{}
}
\startdata
All KOIs & no correction & RV simulation & 0\% & no correction & $3.31\times10^{-16}$ \\
All KOIs & no correction & RV simulation & 0\% & All KOIs & $5.09\times10^{-18}$ \\
All KOIs & no correction & RV simulation & 20\% & All KOIs & $5.56\times10^{-21}$ \\
All KOIs & no correction & RV simulation & 50\% & All KOIs & $1.57\times10^{-14}$ \\
$R_{pl} < 2 R_\oplus$ KOIs & no correction & RV simulation & 0\% & no correction & $3.90\times10^{-23}$ \\
$R_{pl} < 2 R_\oplus$ KOIs & no correction & RV simulation & 0\% & $R_{pl} < 2 R_\oplus$ KOIs & $3.07\times10^{-26}$ \\
$R_{pl} < 2 R_\oplus$ KOIs & no correction & RV simulation & 20\% & $R_{pl} < 2 R_\oplus$ KOIs & $3.53\times10^{-21}$ \\
$R_{pl} < 2 R_\oplus$ KOIs & no correction & RV simulation & 50\% & $R_{pl} < 2 R_\oplus$ KOIs & $1.85\times10^{-19}$ \\
$2 R_{\oplus}<R_{pl}<6 R_{\oplus}$ KOIs & no correction & RV simulation & 0\% & no correction & $3.91\times10^{-12}$ \\
$2 R_{\oplus}<R_{pl}<6 R_{\oplus}$ KOIs & no correction & RV simulation & 0\% & $2 R_{\oplus}<R_{pl}<6 R_{\oplus}$ KOIs & $2.45\times10^{-13}$ \\
$2 R_{\oplus}<R_{pl}<6 R_{\oplus}$ KOIs & no correction & RV simulation & 20\% & $2 R_{\oplus}<R_{pl}<6 R_{\oplus}$ KOIs & $2.11\times10^{-14}$ \\
$2 R_{\oplus}<R_{pl}<6 R_{\oplus}$ KOIs & no correction & RV simulation & 50\% & $2 R_{\oplus}<R_{pl}<6 R_{\oplus}$ KOIs & $2.66\times10^{-15}$ \\
$R_{pl} > 6 R_\oplus$ KOIs & no correction & RV simulation & 0\% & no correction & $4.48\times10^{-4}$ \\
$R_{pl} > 6 R_\oplus$ KOIs & no correction & RV simulation & 0\% & $R_{pl} > 6 R_\oplus$ KOIs & $2.72\times10^{-4}$ \\
$R_{pl} > 6 R_\oplus$ KOIs & no correction & RV simulation & 20\% & $R_{pl} > 6 R_\oplus$ KOIs & $8.63\times10^{-4}$ \\
$R_{pl} > 6 R_\oplus$ KOIs & no correction & RV simulation & 50\% & $R_{pl} > 6 R_\oplus$ KOIs & $2.40\times10^{-2}$ \\
single KOIs & no correction & RV simulation & 0\% & no correction & $2.90\times10^{-15}$ \\
single KOIs & no correction & RV simulation & 0\% & single KOIs & $4.80\times10^{-17}$ \\
single KOIs & no correction & RV simulation & 20\% & single KOIs & $3.86\times10^{-13}$ \\
single KOIs & no correction & RV simulation & 50\% & single KOIs & $3.92\times10^{-15}$ \\
multi KOIs & no correction & RV simulation & 0\% & no correction & $2.54\times10^{-16}$ \\
multi KOIs & no correction & RV simulation & 0\% & multi KOIs & $4.76\times10^{-18}$ \\
multi KOIs & no correction & RV simulation & 20\% & multi KOIs & $6.40\times10^{-22}$ \\
multi KOIs & no correction & RV simulation & 50\% & multi KOIs & $2.42\times10^{-11}$ \\
\hline
All KOIs & no correction & $R_{pl} < 2 R_\oplus$ KOIs & 0\% & no correction & $1.76\times10^{-5}$ \\
All KOIs & no correction & $2 R_{\oplus}<R_{pl}<6 R_{\oplus}$ KOIs & 0\% & no correction & $1.67\times10^{-1}$ \\
All KOIs & no correction & $R_{pl} > 6 R_\oplus$ KOIs & 0\% & no correction & $4.60\times10^{-10}$ \\
All KOIs & no correction & single KOIs & 0\% & no correction & $1.23\times10^{-1}$ \\
All KOIs & no correction & multi KOIs & 0\% & no correction & $1.13\times10^{-2}$ \\
$R_{pl} < 2 R_\oplus$ KOIs & no correction & $2 R_{\oplus}<R_{pl}<6 R_{\oplus}$ KOIs & 0\% & no correction & $4.19\times10^{-8}$ \\
$R_{pl} < 2 R_\oplus$ KOIs & no correction & $R_{pl} > 6 R_\oplus$ KOIs & 0\% & no correction & $5.34\times10^{-17}$ \\
$R_{pl} < 2 R_\oplus$ KOIs & no correction & single KOIs & 0\% & no correction & $1.38\times10^{-8}$ \\
$R_{pl} < 2 R_\oplus$ KOIs & no correction & multi KOIs & 0\% & no correction & $2.28\times10^{-4}$ \\
$2 R_{\oplus}<R_{pl}<6 R_{\oplus}$ KOIs & no correction & $R_{pl} > 6 R_\oplus$ KOIs & 0\% & no correction & $4.62\times10^{-8}$ \\
$2 R_{\oplus}<R_{pl}<6 R_{\oplus}$ KOIs & no correction & single KOIs & 0\% & no correction & $2.94\times10^{-2}$ \\
$2 R_{\oplus}<R_{pl}<6 R_{\oplus}$ KOIs & no correction & multi KOIs & 0\% & no correction & $6.30\times10^{-4}$ \\
$R_{pl} > 6 R_\oplus$ KOIs & no correction & single KOIs & 0\% & no correction & $3.59\times10^{-7}$ \\
$R_{pl} > 6 R_\oplus$ KOIs & no correction & multi KOIs & 0\% & no correction & $2.94\times10^{-13}$ \\
single KOIs & no correction & multi KOIs & 0\% & no correction & $1.66\times10^{-5}$ \\
\hline
All KOIs & no correction & $R_{pl} < 2 R_\oplus$ KOIs & 0\% & All KOIs & $9.55\times10^{-8}$ \\
All KOIs & no correction & $2 R_{\oplus}<R_{pl}<6 R_{\oplus}$ KOIs & 0\% & All KOIs & $1.85\times10^{-3}$ \\
All KOIs & no correction & $R_{pl} > 6 R_\oplus$ KOIs & 0\% & All KOIs & $1.27\times10^{-13}$ \\
All KOIs & no correction & single KOIs & 0\% & All KOIs & $1.75\times10^{-1}$ \\
All KOIs & no correction & multi KOIs & 0\% & All KOIs & $4.20\times10^{-2}$ \\
$R_{pl} < 2 R_\oplus$ KOIs & All KOIs & $2 R_{\oplus}<R_{pl}<6 R_{\oplus}$ KOIs & 0\% & All KOIs & $5.08\times10^{-14}$ \\
$R_{pl} < 2 R_\oplus$ KOIs & All KOIs & $R_{pl} > 6 R_\oplus$ KOIs & 0\% & All KOIs & $5.21\times10^{-21}$ \\
$R_{pl} < 2 R_\oplus$ KOIs & All KOIs & single KOIs & 0\% & All KOIs & $5.17\times10^{-10}$ \\
$R_{pl} < 2 R_\oplus$ KOIs & All KOIs & multi KOIs & 0\% & All KOIs & $6.52\times10^{-8}$ \\
$2 R_{\oplus}<R_{pl}<6 R_{\oplus}$ KOIs & All KOIs & $R_{pl} > 6 R_\oplus$ KOIs & 0\% & All KOIs & $5.51\times10^{-10}$ \\
$2 R_{\oplus}<R_{pl}<6 R_{\oplus}$ KOIs & All KOIs & single KOIs & 0\% & All KOIs & $5.92\times10^{-3}$ \\
$2 R_{\oplus}<R_{pl}<6 R_{\oplus}$ KOIs & All KOIs & multi KOIs & 0\% & All KOIs & $5.89\times10^{-6}$ \\
$R_{pl} > 6 R_\oplus$ KOIs & All KOIs & single KOIs & 0\% & All KOIs & $9.71\times10^{-12}$ \\
$R_{pl} > 6 R_\oplus$ KOIs & All KOIs & multi KOIs & 0\% & All KOIs & $6.26\times10^{-15}$ \\
single KOIs & All KOIs & multi KOIs & 0\% & All KOIs & $2.40\times10^{-4}$ \\
\enddata
\tablenotetext{a}{No additional random errors added to sample one as listed in this table.}
\end{deluxetable}

\begin{deluxetable}{lll}
\tabletypesize{\scriptsize}
\tablewidth{0pt}
\tablecolumns{3}
\tablecaption{Linear fits to median trends in $\alpha$ in Figures 4--6}
\tablehead{
\colhead{KOI List} & \colhead{Slope} & \colhead{Significance ($\sigma$)}}
\startdata
\multicolumn{3}{c}{Stellar Radius (R$_\odot^{-1}$)} \\
B13 All KOIs & -0.27$\pm$0.05 & 5.4 \\
\citet[]{borucki3} & -0.33$\pm$0.05 & 6.6\\ 
B13 single KOIs & -0.26$\pm$0.06 & 4.3 \\
B13 multi KOIs & -0.28$\pm$0.06 & 4.7 \\
\hline
\multicolumn{3}{c}{Stellar Mass (M$_\odot^{-1})$} \\
B13 All KOIs & -0.28$\pm$0.09 & 3.1 \\
\citet[]{borucki3} & -0.28$\pm$0.09 & 3.1 \\
B13 single KOIs & -0.23$\pm$0.10 & 2.3 \\
B13 multi KOIs & -0.23$\pm$0.09 & 2.6 \\
\hline
\multicolumn{3}{c}{Stellar Temp (K$^{-1})$}\\
B13 All KOIs & -(3.2$\pm$2.6)$\times10^{-5}$ & 0.8 \\
\citet[]{borucki3} & -(8.3$\pm$4.7)$\times10^{-5}$ & 1.8 \\
B13 single KOIs & -(1.7$\pm$2.6)$\times10^{-5}$ & 0.7 \\
B13 multi KOIs & -(5.0$\pm$4.1)$\times10^{-5}$ & 1.2 \\

\enddata
\end{deluxetable}

\begin{figure}
\centering
\includegraphics[width=0.8\textwidth,clip=true,trim=0cm 0cm 0cm 0cm]{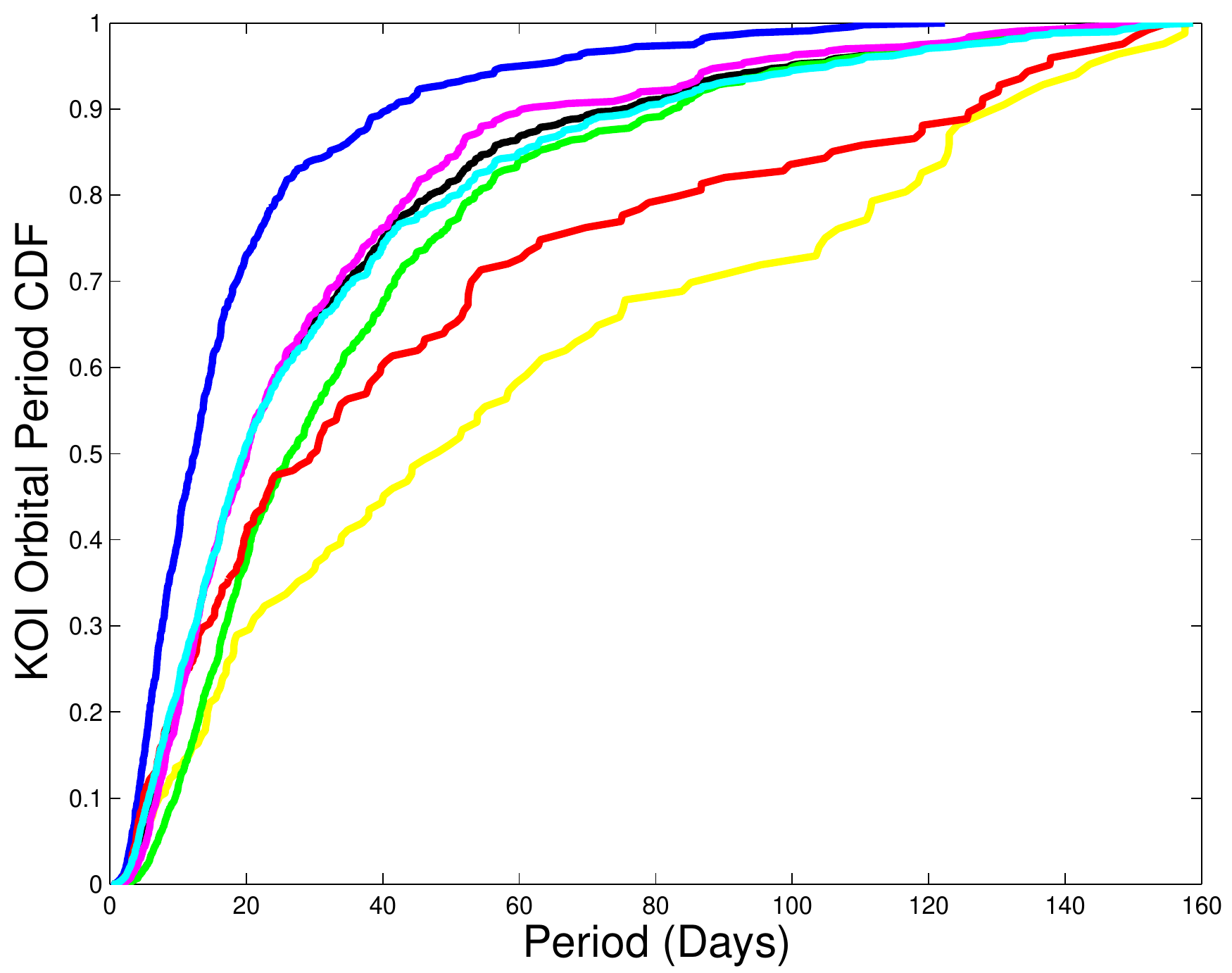}  
\caption{CDFs for orbital periods for the RV sample, KOIs and subsets thereof.  KOIs with $R_{pl}<2 R_{\oplus}$, $2 R_{\oplus}<R_{pl}<6 R_{\oplus}$, $R_{pl}>6 R_{\oplus}$, singles, multis and all are shown in blue, green, red, cyan, magenta and black respectively.  The CDF for RV planet orbital periods is shown in yellow.}
\end{figure}

\begin{figure}
\centering
\includegraphics[scale=0.35,clip=true,trim=0cm 0cm 0cm 0cm]{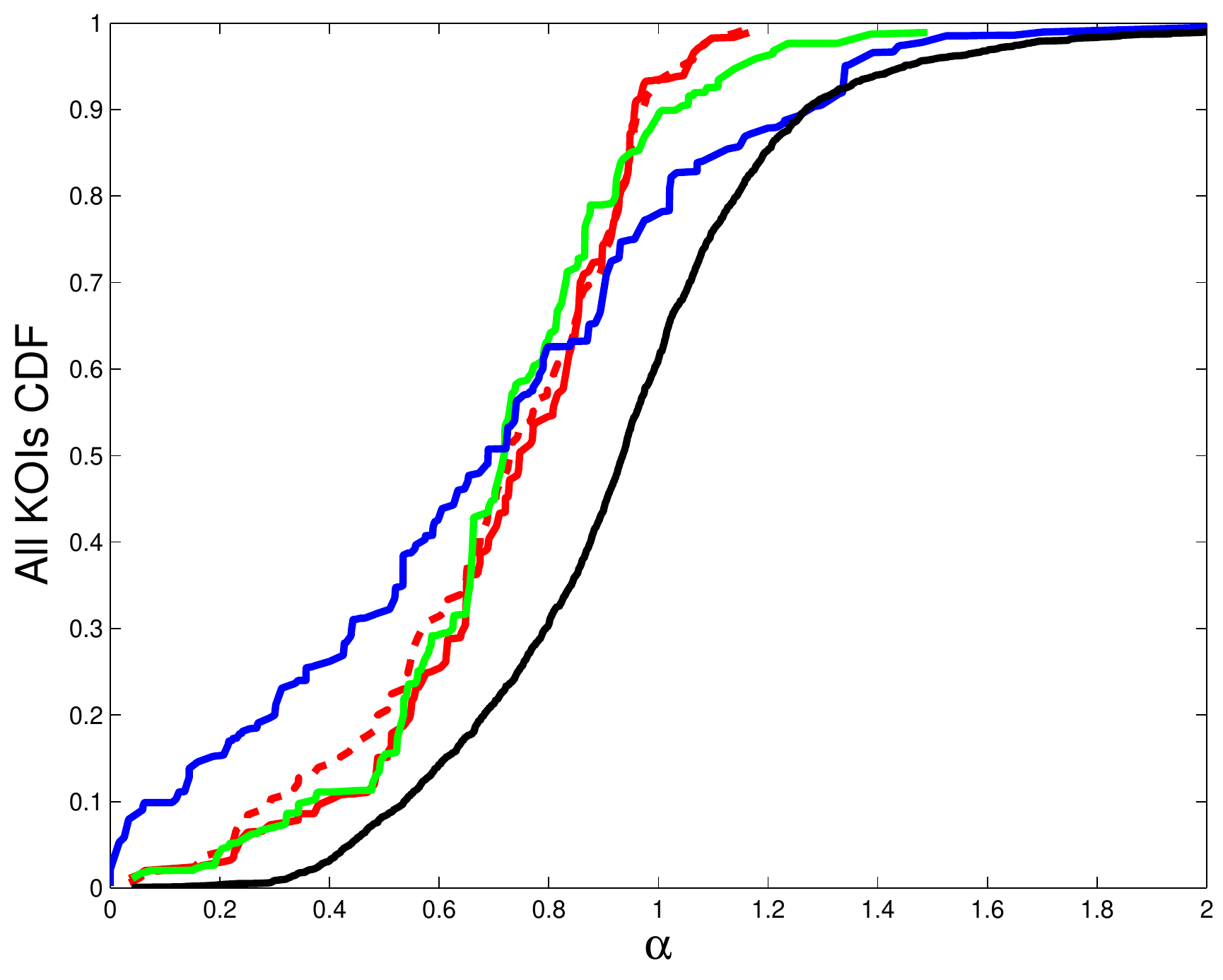}  
\includegraphics[scale=0.35,clip=true,trim=0cm 0cm 0cm 0cm]{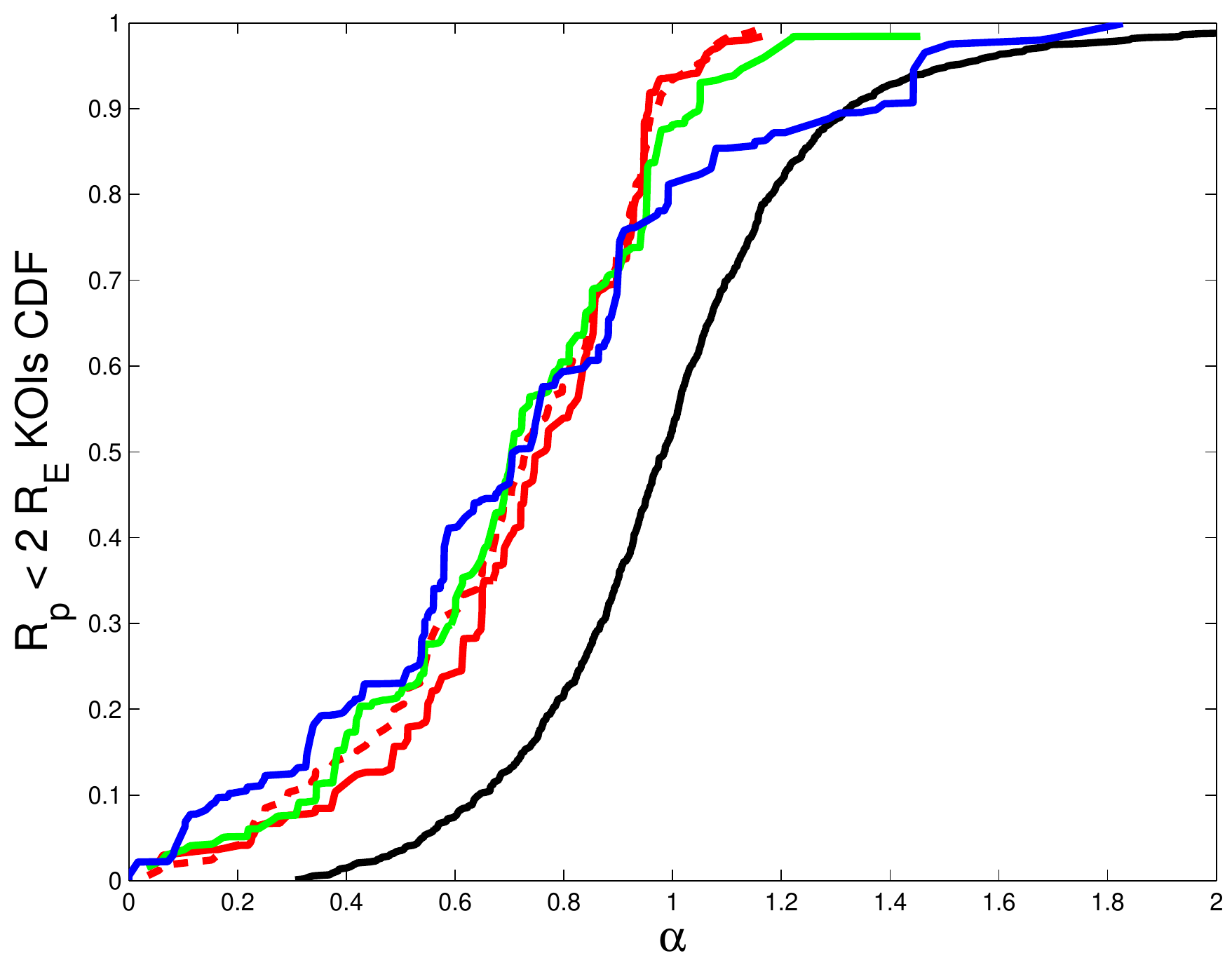}  \\
\includegraphics[scale=0.35,clip=true,trim=0cm 0cm 0cm 0cm]{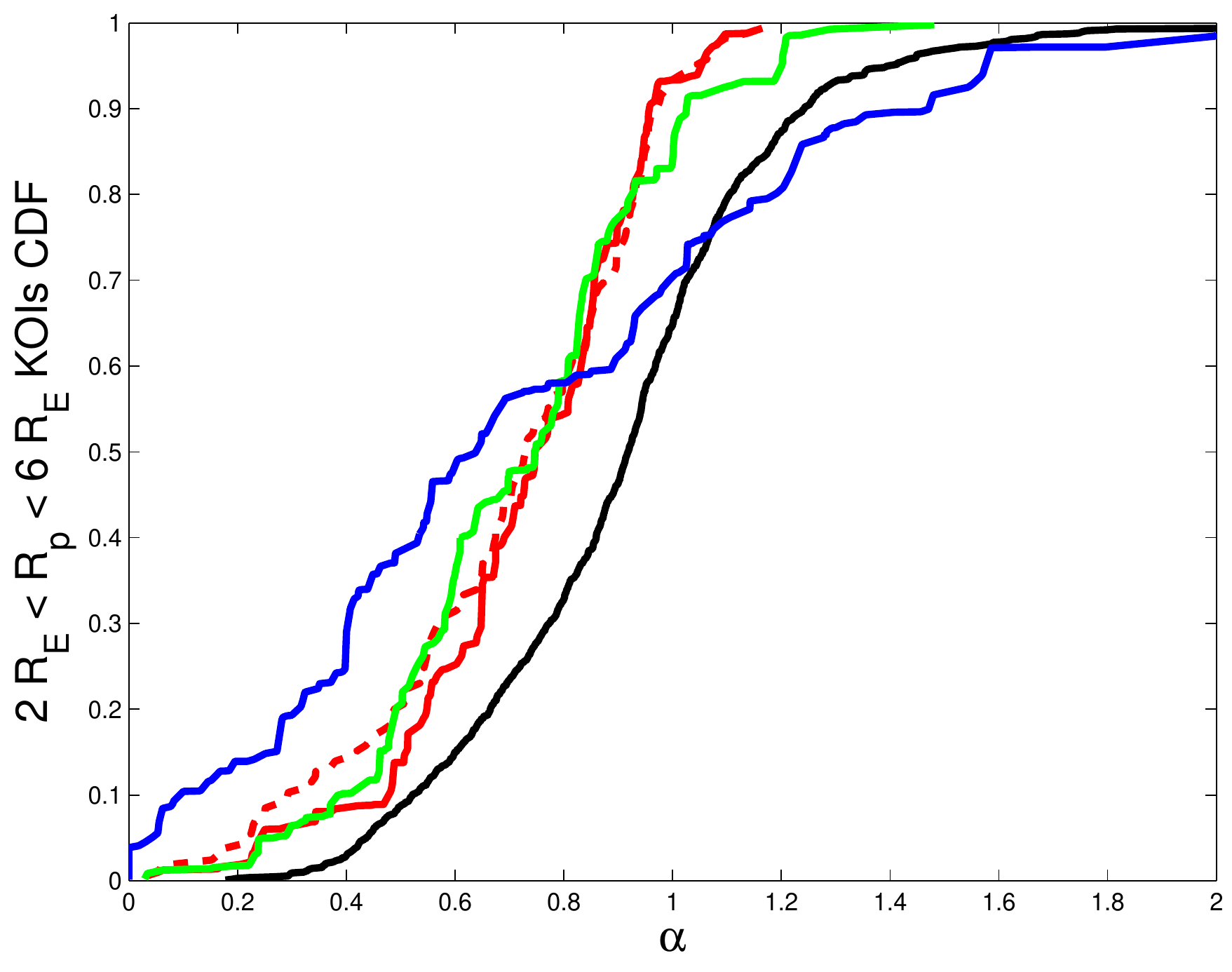}  
\includegraphics[scale=0.35,clip=true,trim=0cm 0cm 0cm 0cm]{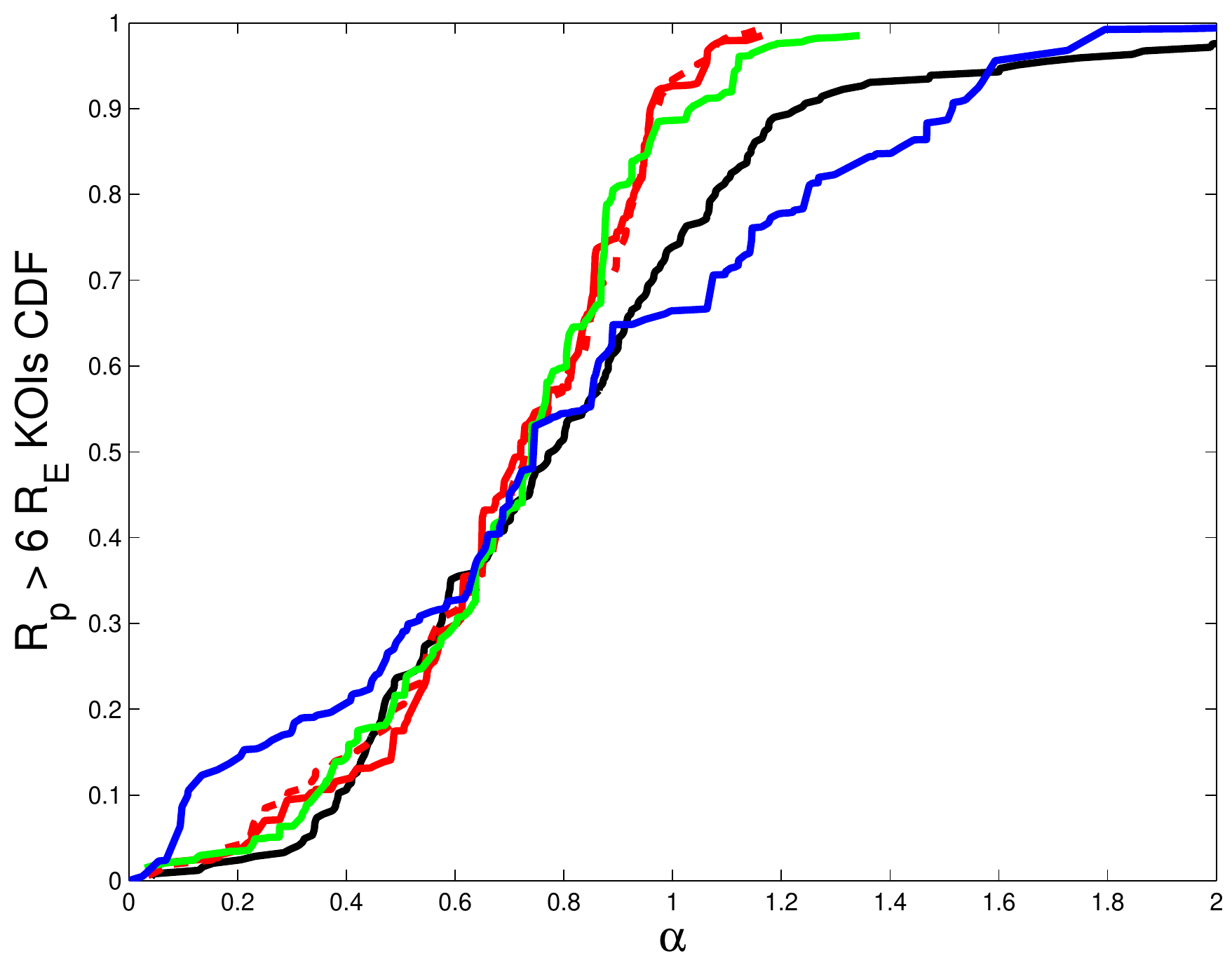} \\  
\includegraphics[scale=0.35,clip=true,trim=0cm 0cm 0cm 0cm]{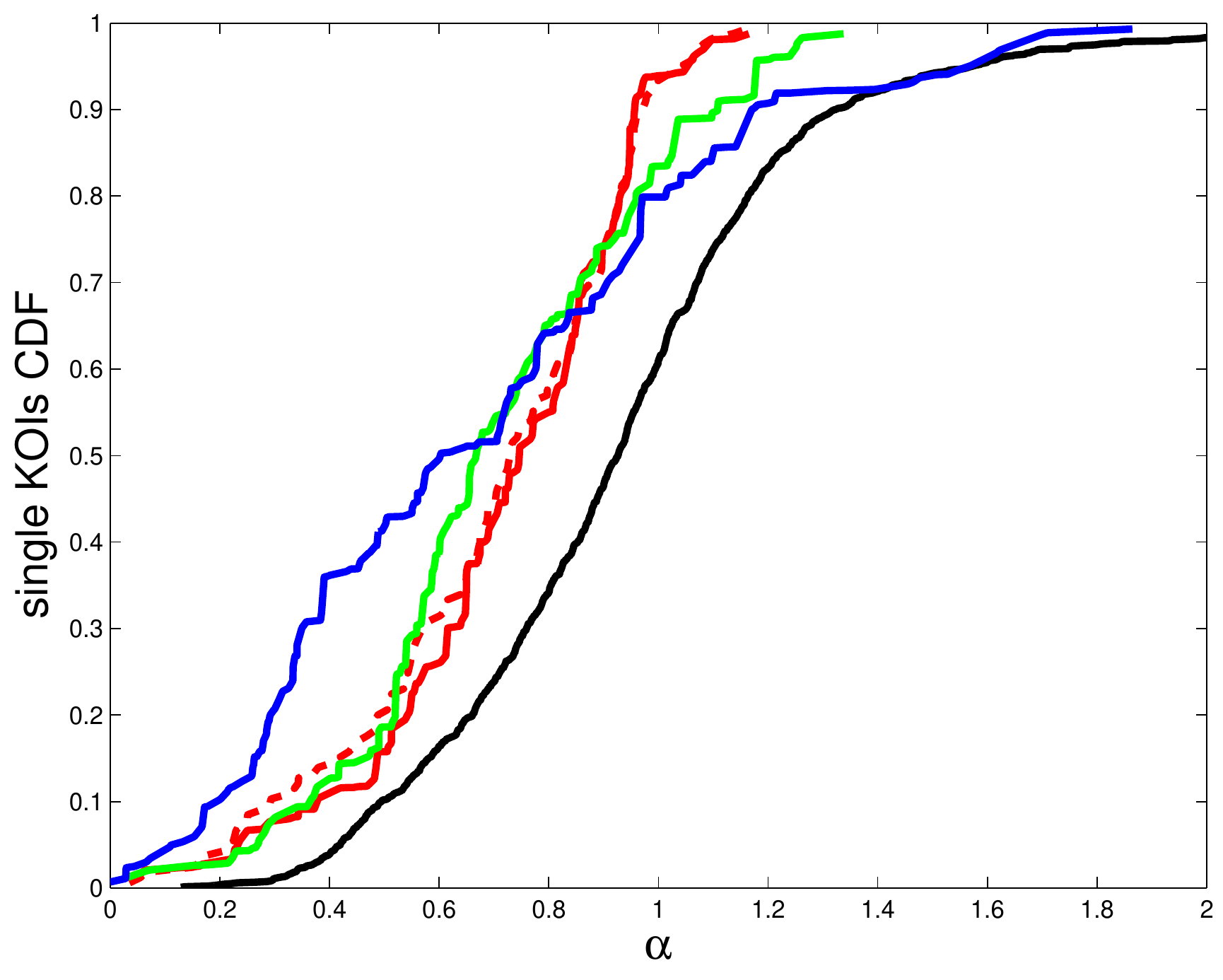}  
\includegraphics[scale=0.35,clip=true,trim=0cm 0cm 0cm 0cm]{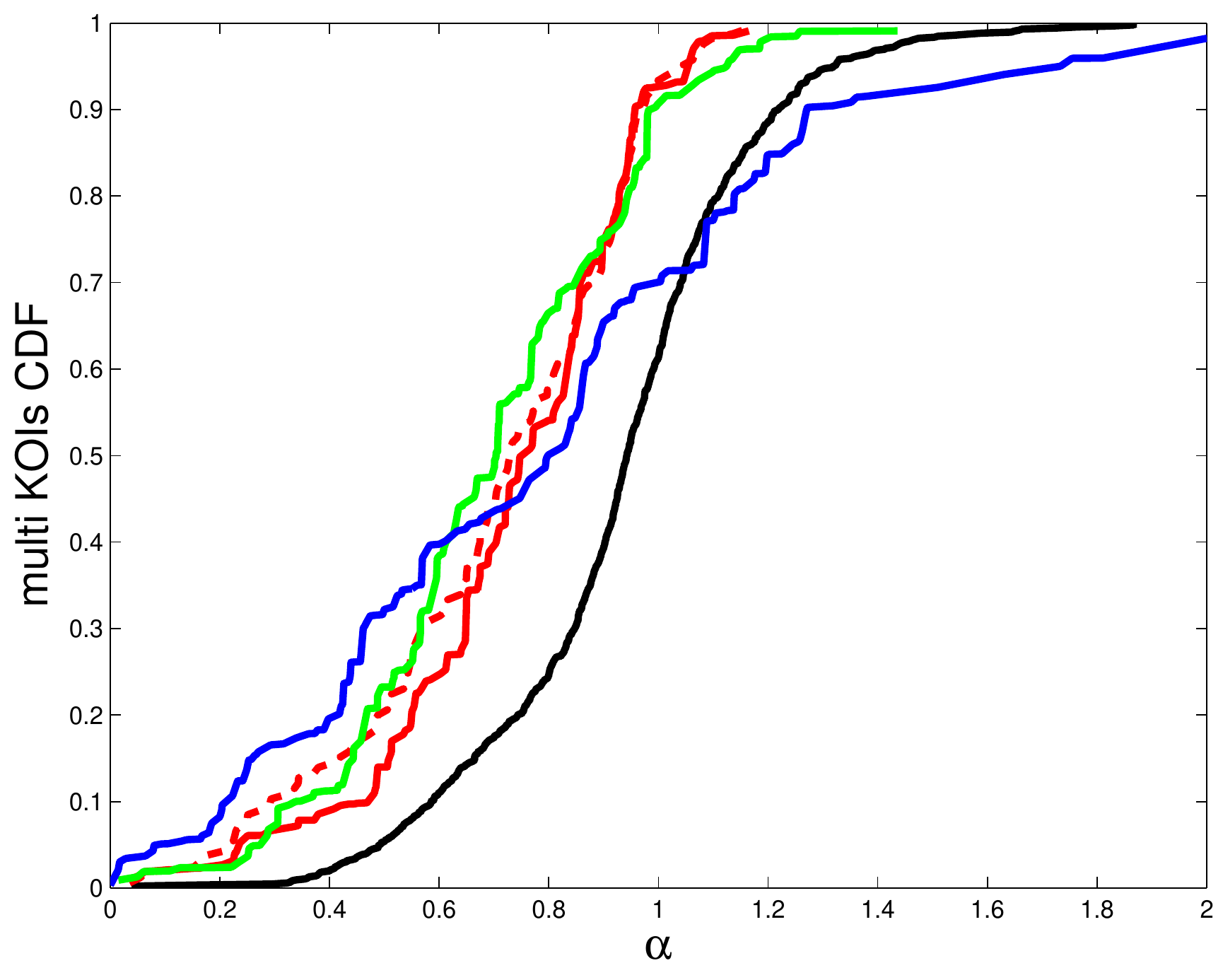} \\  

\caption{Black curves: CDFs as a function of transit duration anomalies $\alpha$ for all KOIs from B13 in the top left, $R_{pl}<2 R_{\oplus}$ KOIs in the top right, $2 R_{\oplus}<R_{pl}<6 R_{\oplus}$ KOIs in the middle left,  $R_{pl}>6 R_{\oplus}$ KOIs in the middle right, single KOIs in the bottom left, and multi KOIs in the bottom right.   Red-dashed curves, identical in all panels: Simulated CDF($\alpha$) from RV-discovered exoplanets with no corrections.  Red curves, all panels: Simulated CDF($\alpha$) from RV-discovered exoplanets, corrected (weighted) to match the period distribution of the KOIs or appropriate subset thereof.  Green curves, all panels: Same as the red curves, with 20\% additional random errors added in $\alpha$ to represent stellar parameter errors.  Blue curves, all panels: Same as the red curves, with 50\% additional random errors added in $\alpha$.  See ${\S}$5 for discussion.}
\end{figure}

\begin{figure}
\centering
\includegraphics[width=0.8\textwidth,clip=true,trim=0.93cm 0.77cm 0cm 0cm]{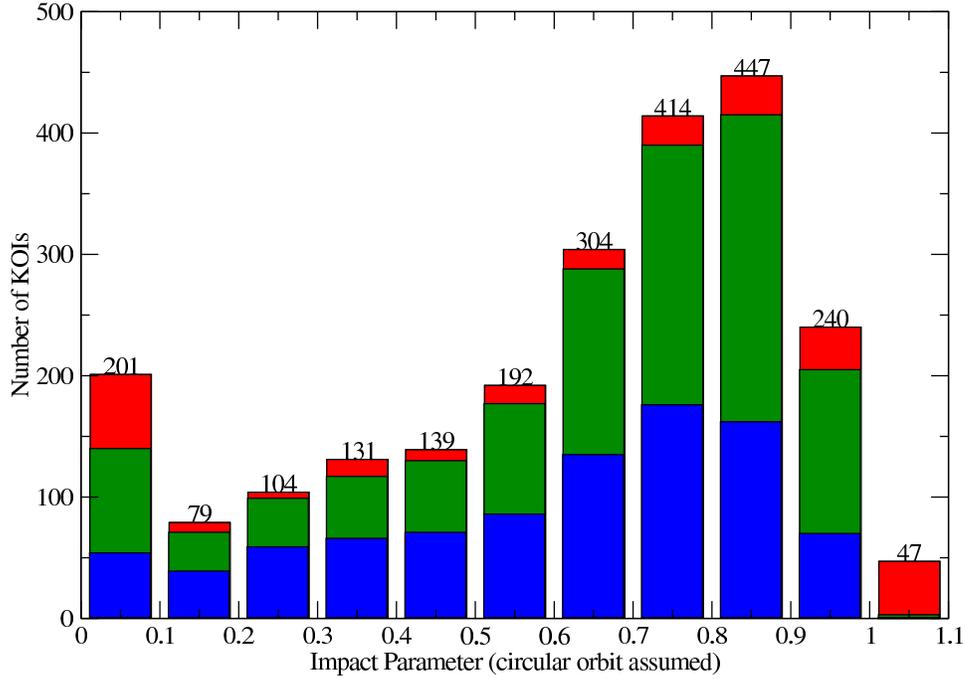}  
\caption{Impact parameter cumulative histogram for all KOIs, assuming circular orbits (B13).  The contributions of KOIs with $R_{pl}<2 R_{\oplus}$, $2 R_{\oplus}<R_{pl}<6 R_{\oplus}$, and $R_{pl}>6 R_{\oplus}$ are shown in blue, green and red respectively.}
\end{figure}

\begin{figure}
\centering
\includegraphics[scale=0.45,clip=true,trim=2cm 12cm 2cm 3cm]{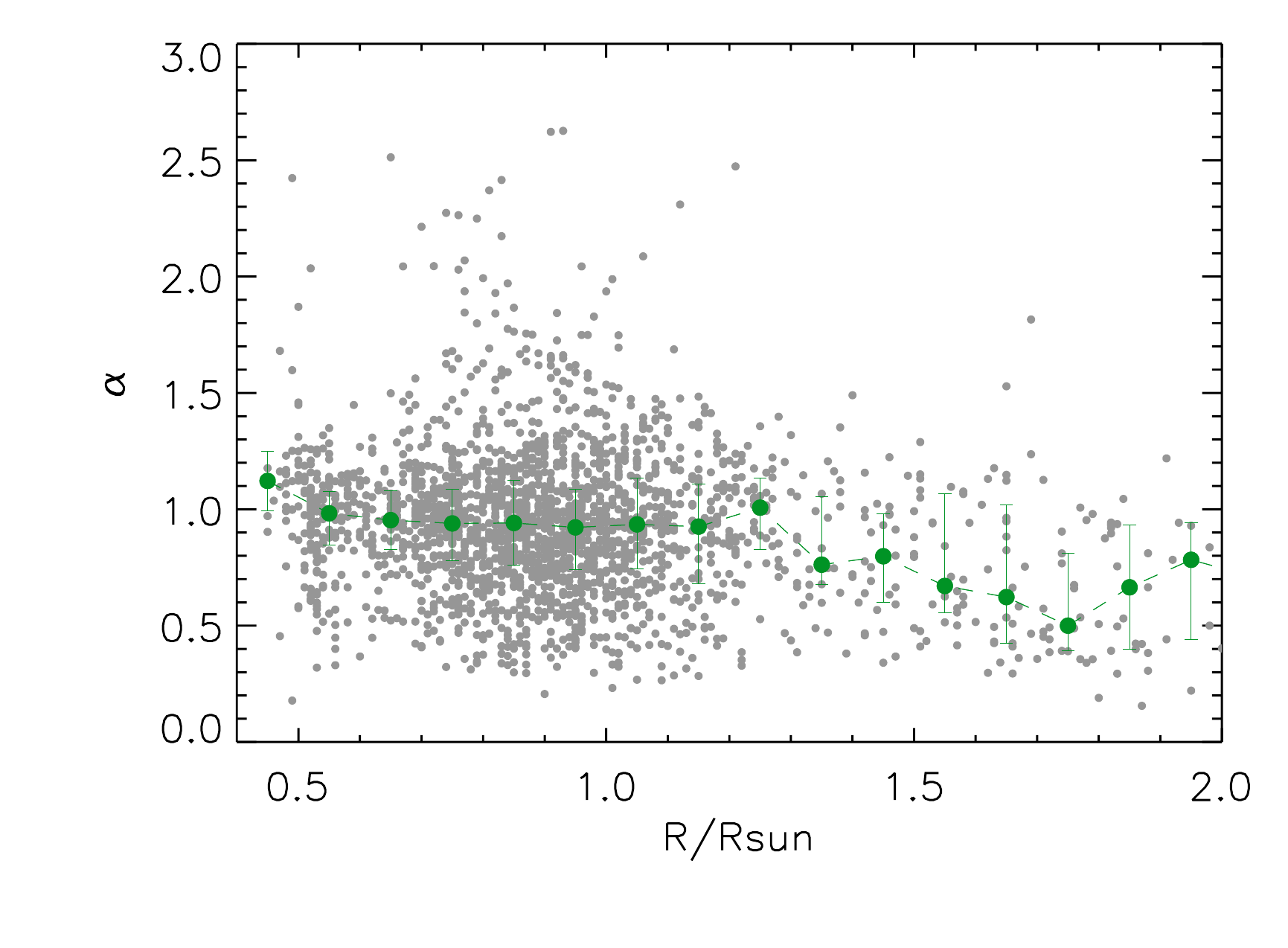}  
\includegraphics[scale=0.45,clip=true,trim=2cm 12cm 2cm 3cm]{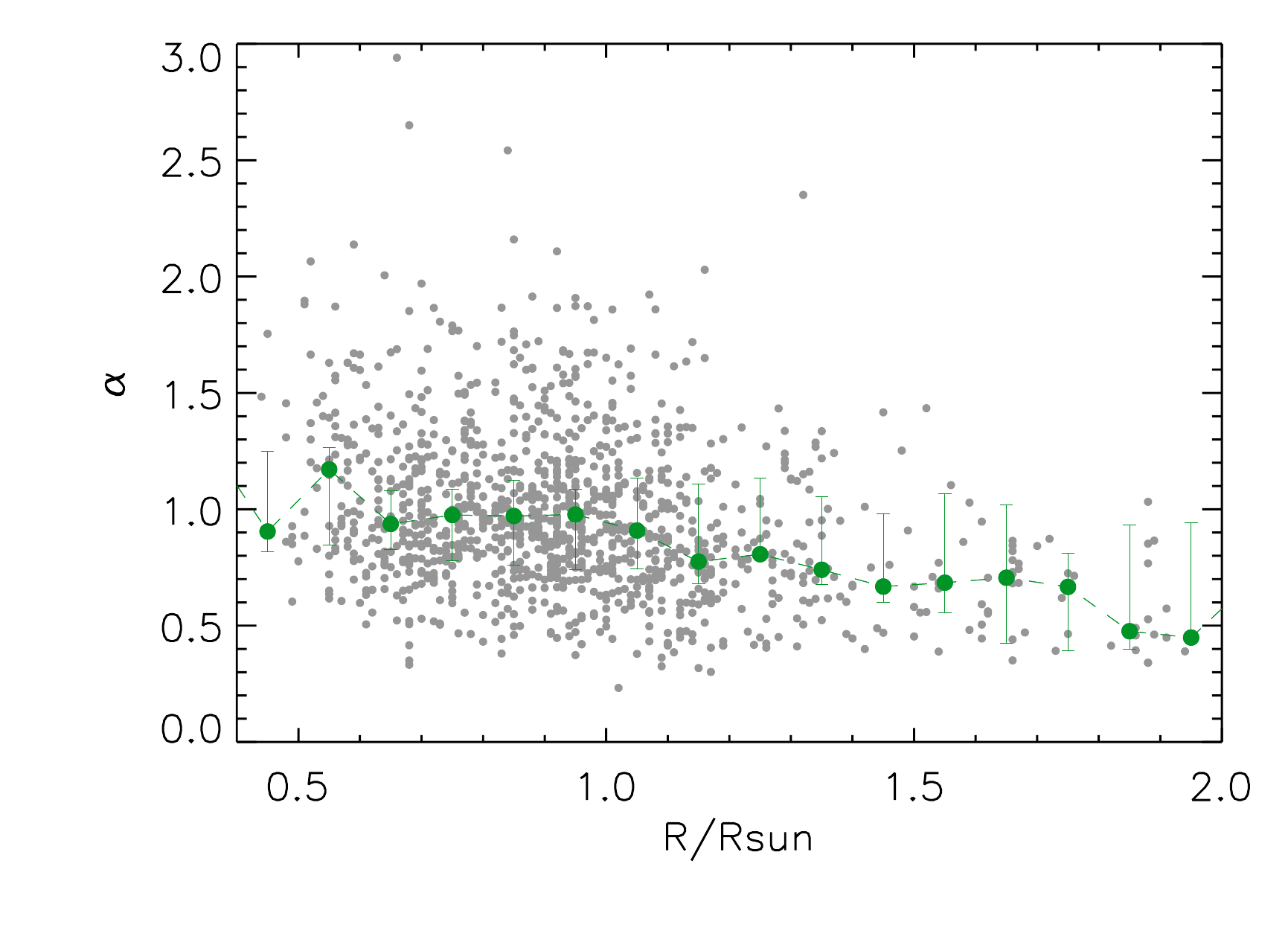}  \\
\includegraphics[scale=0.45,clip=true,trim=0cm 0cm 0cm 0cm]{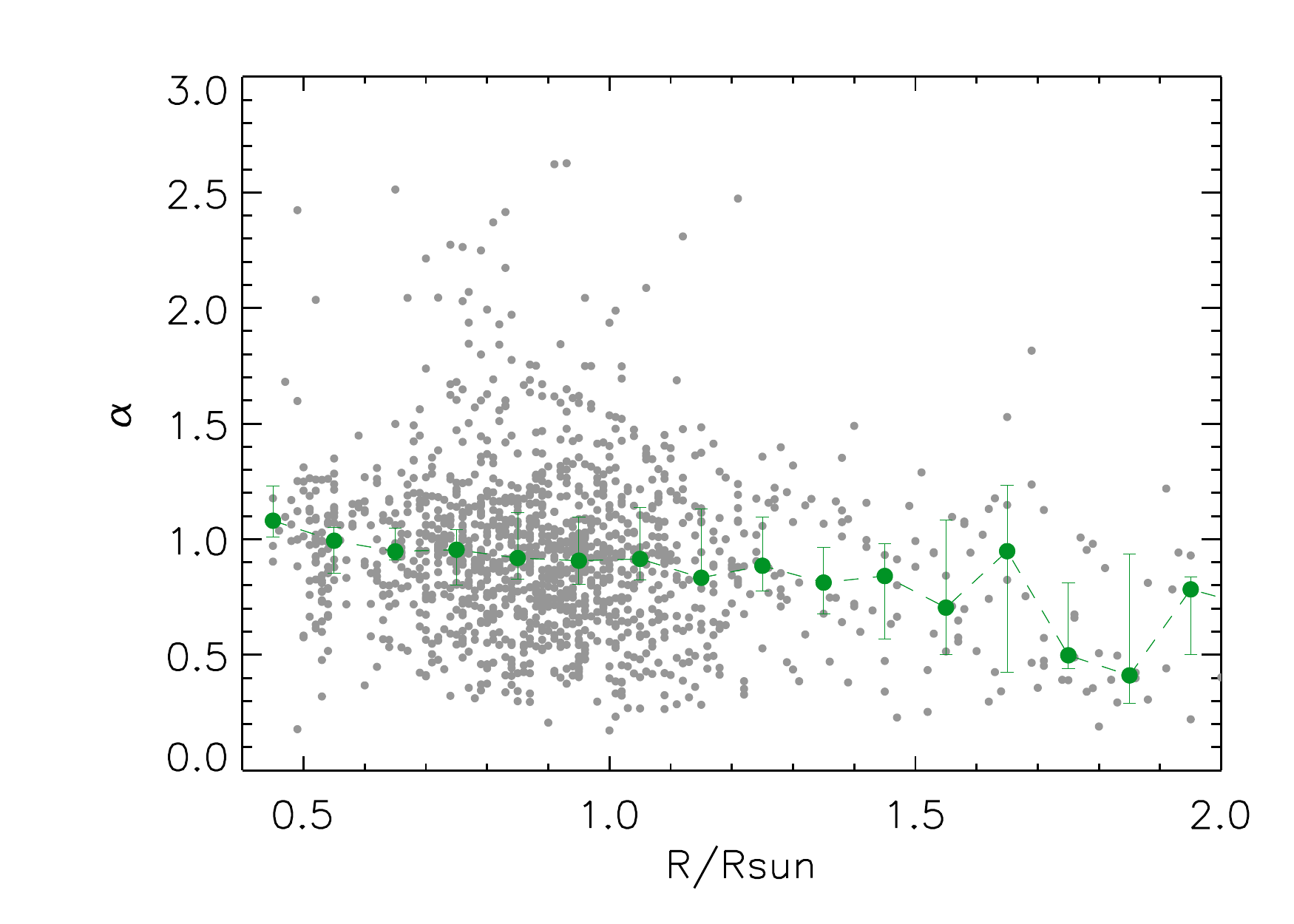}  
\includegraphics[scale=0.45,clip=true,trim=0cm 0cm 0cm 0cm]{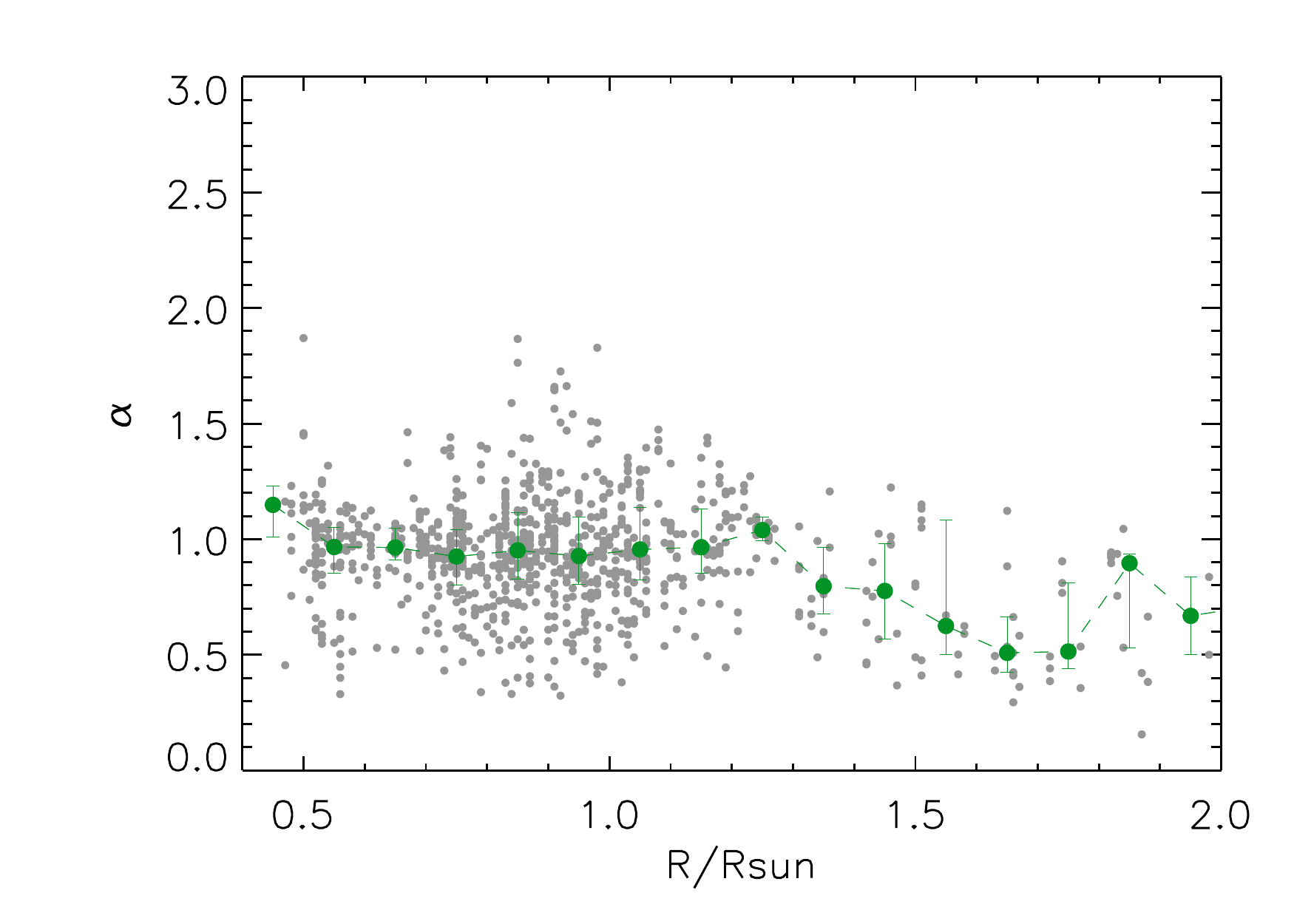}  

\caption{Distributions for the transit duration anomalies $\alpha$ as a function of stellar radius for the KOI list from B13 in the top left, the KOI list from \citet[]{borucki3} in the top right, single KOIs from B13 in the bottom left, and multis from B13 in the bottom right, respectively.  Overplotted are binned median, lower and upper quartile ranges.  Linear fits to the binned median values (not shown) yield slope coefficients in Table 5 that indicating statistically significant ensemble trends in the transit duration (and implied average eccentricity) as a function of stellar radius and mass but not temperature.}  
\end{figure}

\begin{figure}
\centering
\includegraphics[scale=0.45,clip=true,trim=2cm 12cm 2cm 3cm]{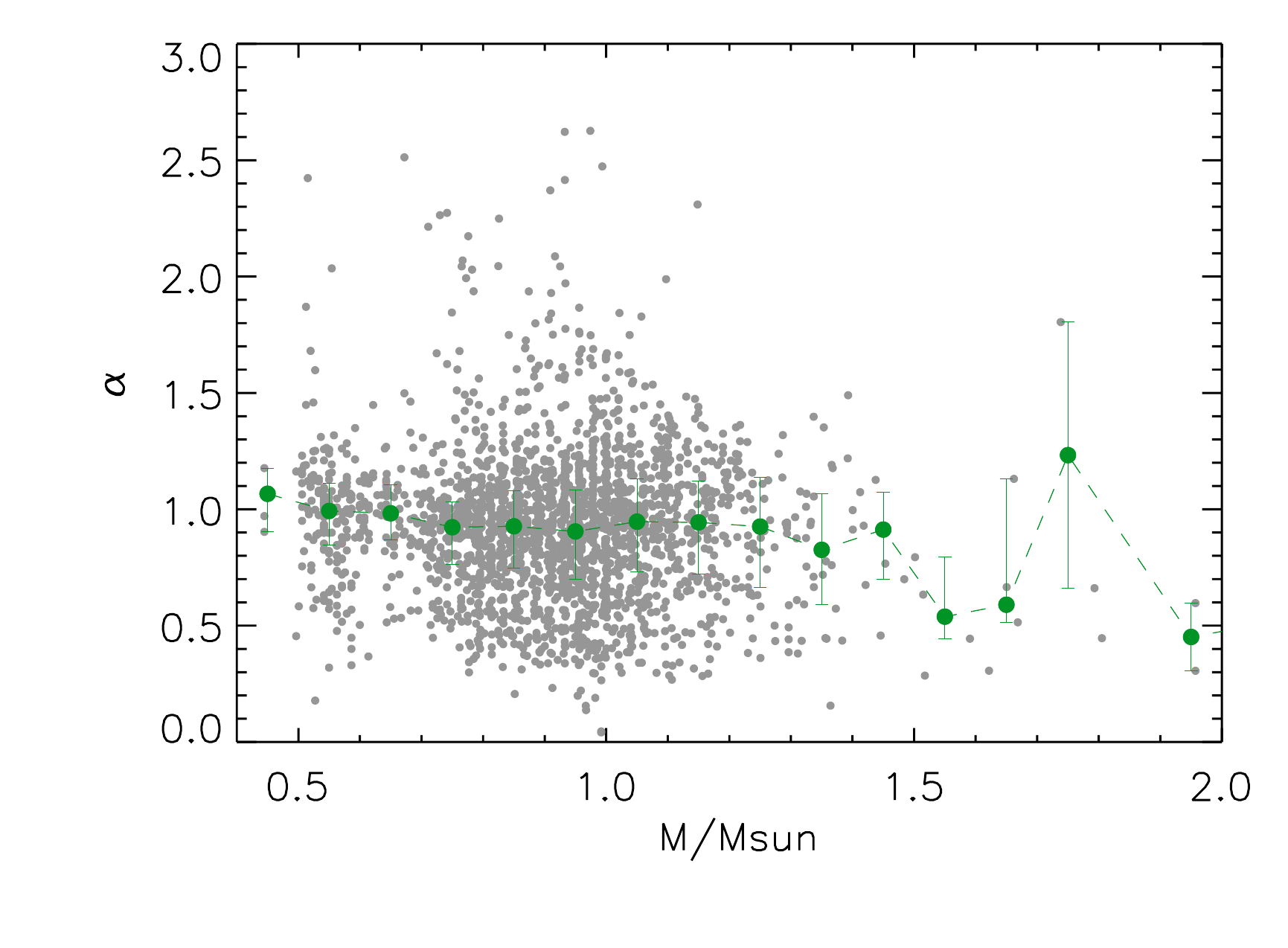}  
\includegraphics[scale=0.45,clip=true,trim=2cm 12cm 2cm 3cm]{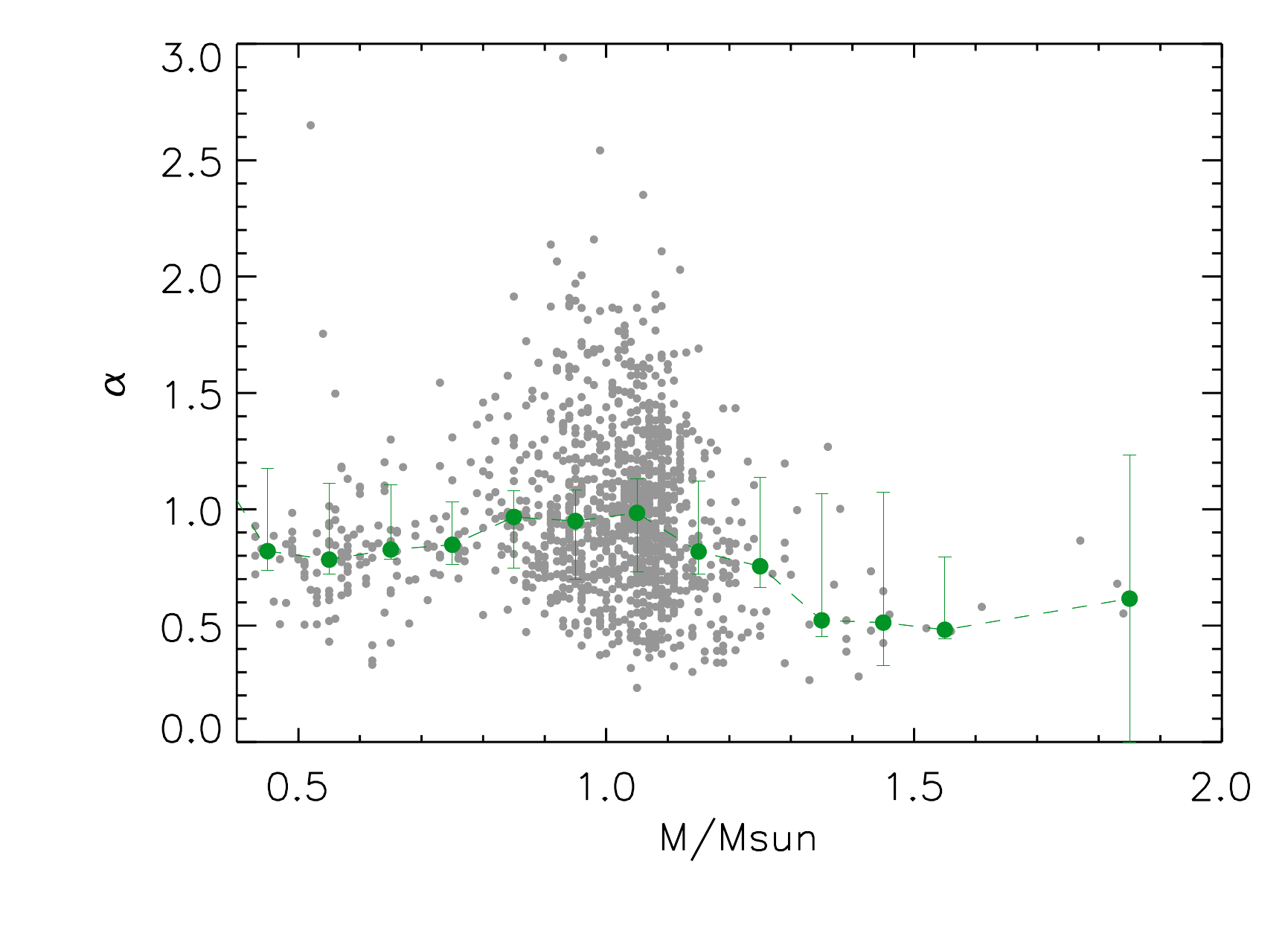}  \\
\includegraphics[scale=0.45,clip=true,trim=0cm 0cm 0cm 0cm]{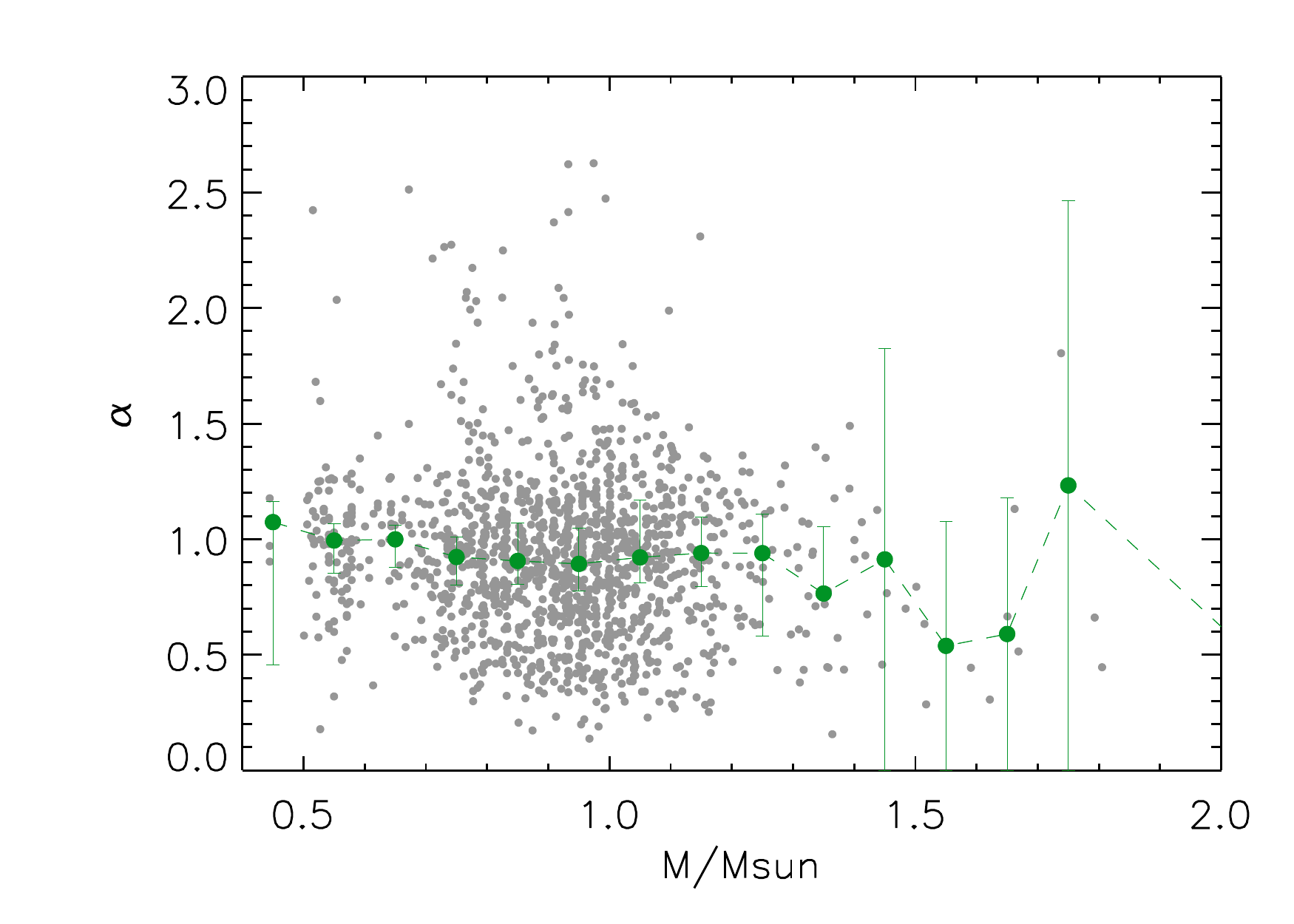}  
\includegraphics[scale=0.45,clip=true,trim=0cm 0cm 0cm 0cm]{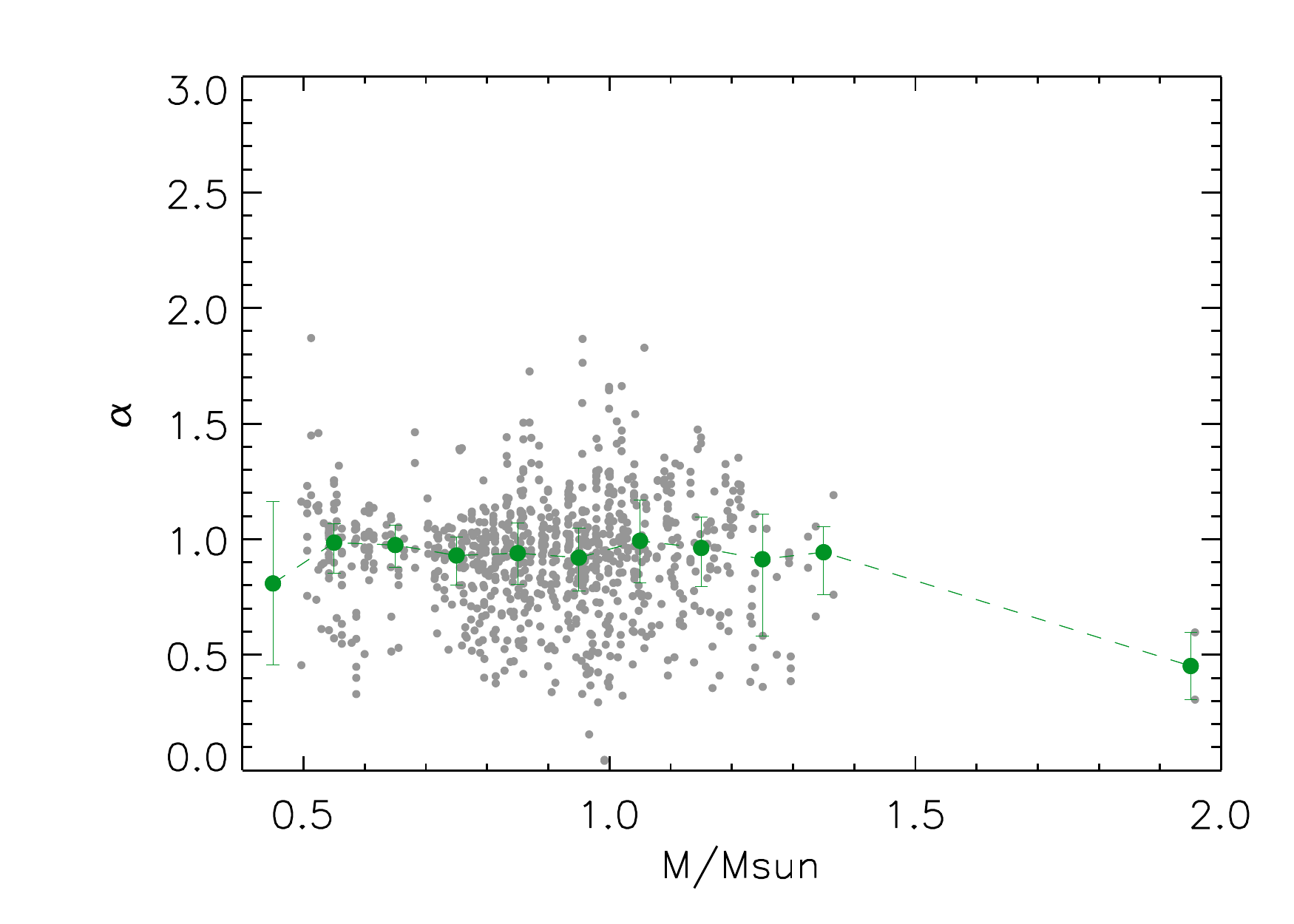}  
\caption{Same as Figure 4 for stellar mass.}
\end{figure}

\begin{figure}
\centering
\includegraphics[scale=0.45,clip=true,trim=2cm 12cm 2cm 3cm]{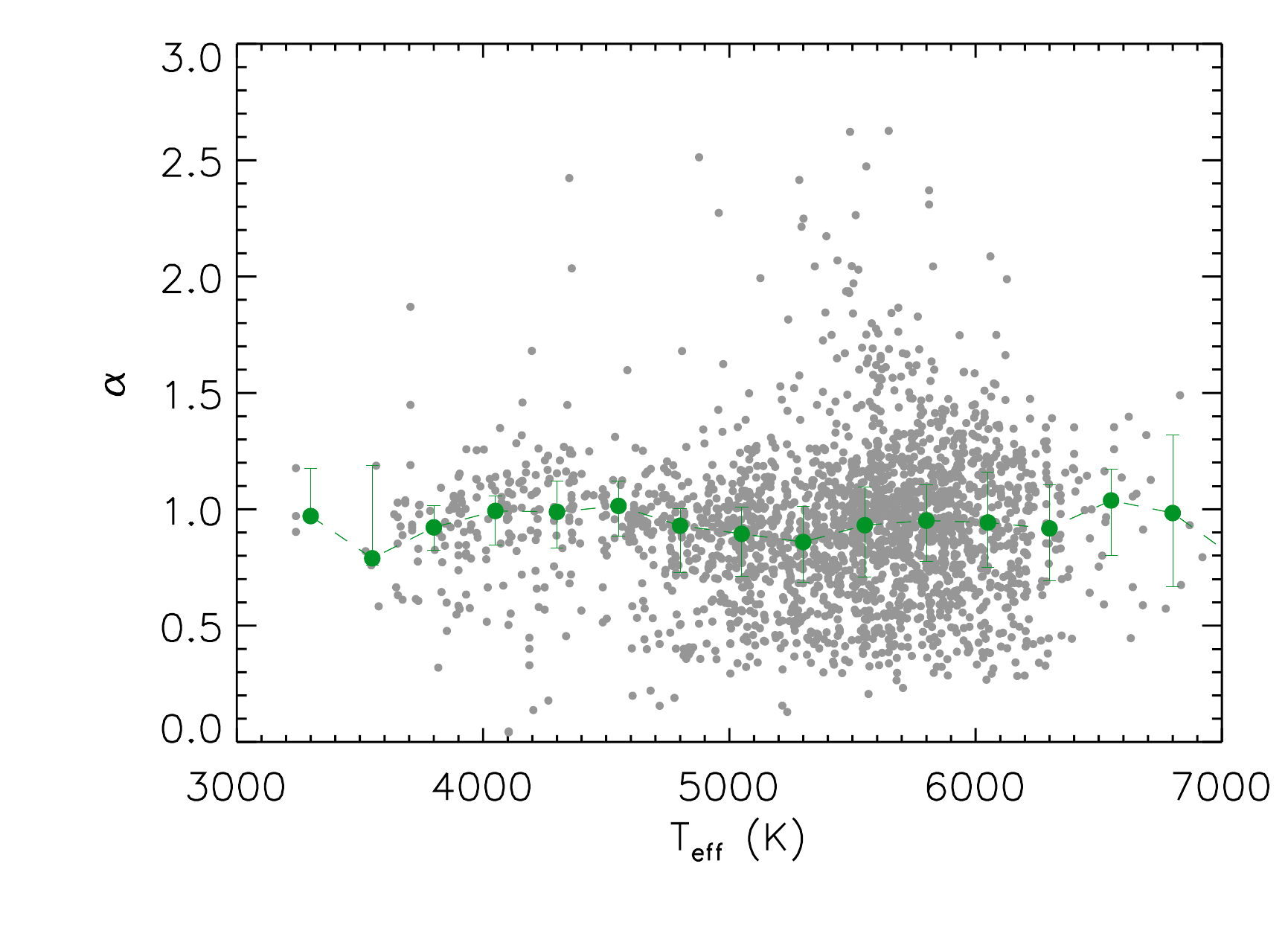}  
\includegraphics[scale=0.45,clip=true,trim=2cm 12cm 2cm 3cm]{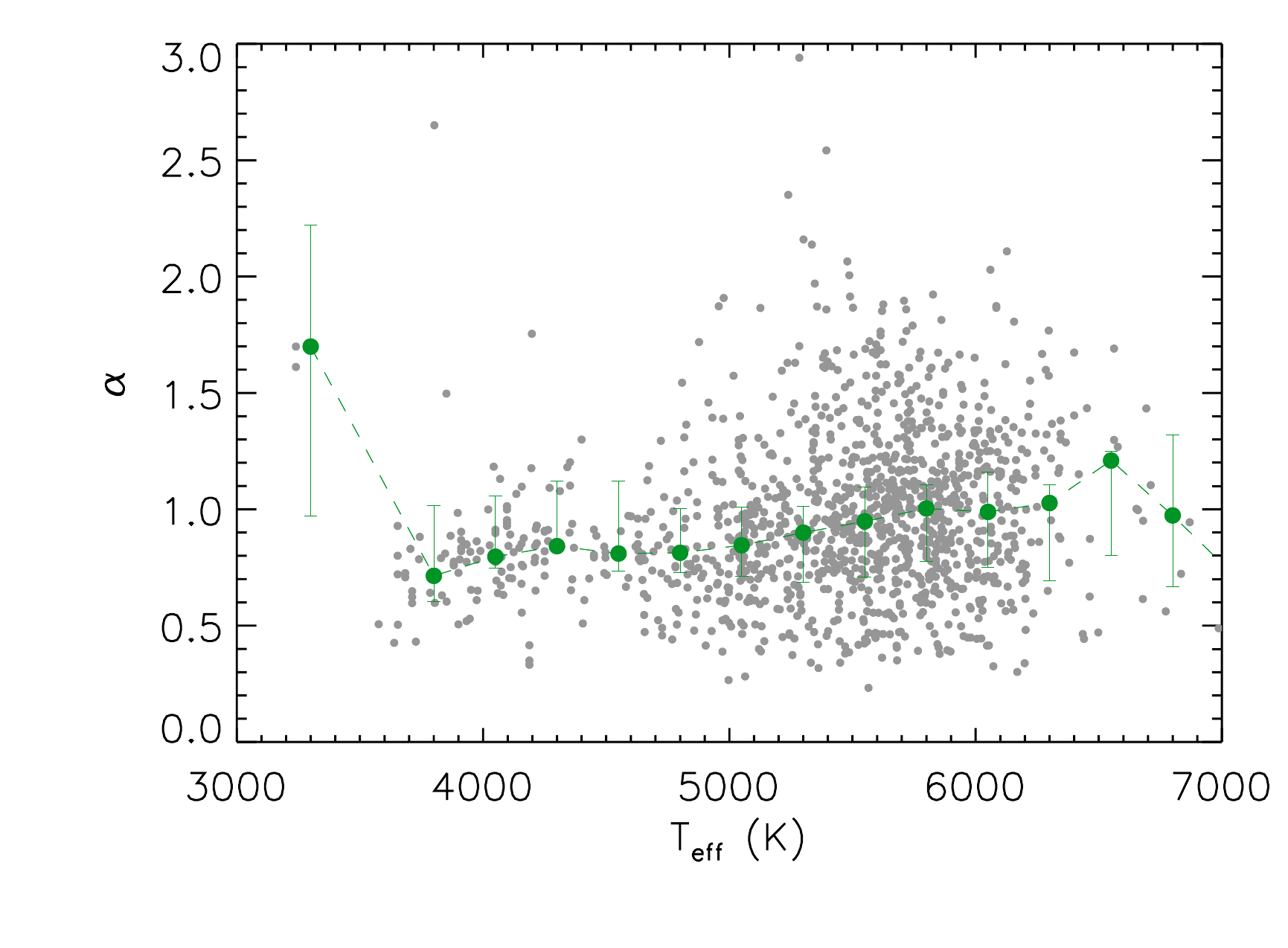}  \\
\includegraphics[scale=0.45,clip=true,trim=0cm 0cm 0cm 0cm]{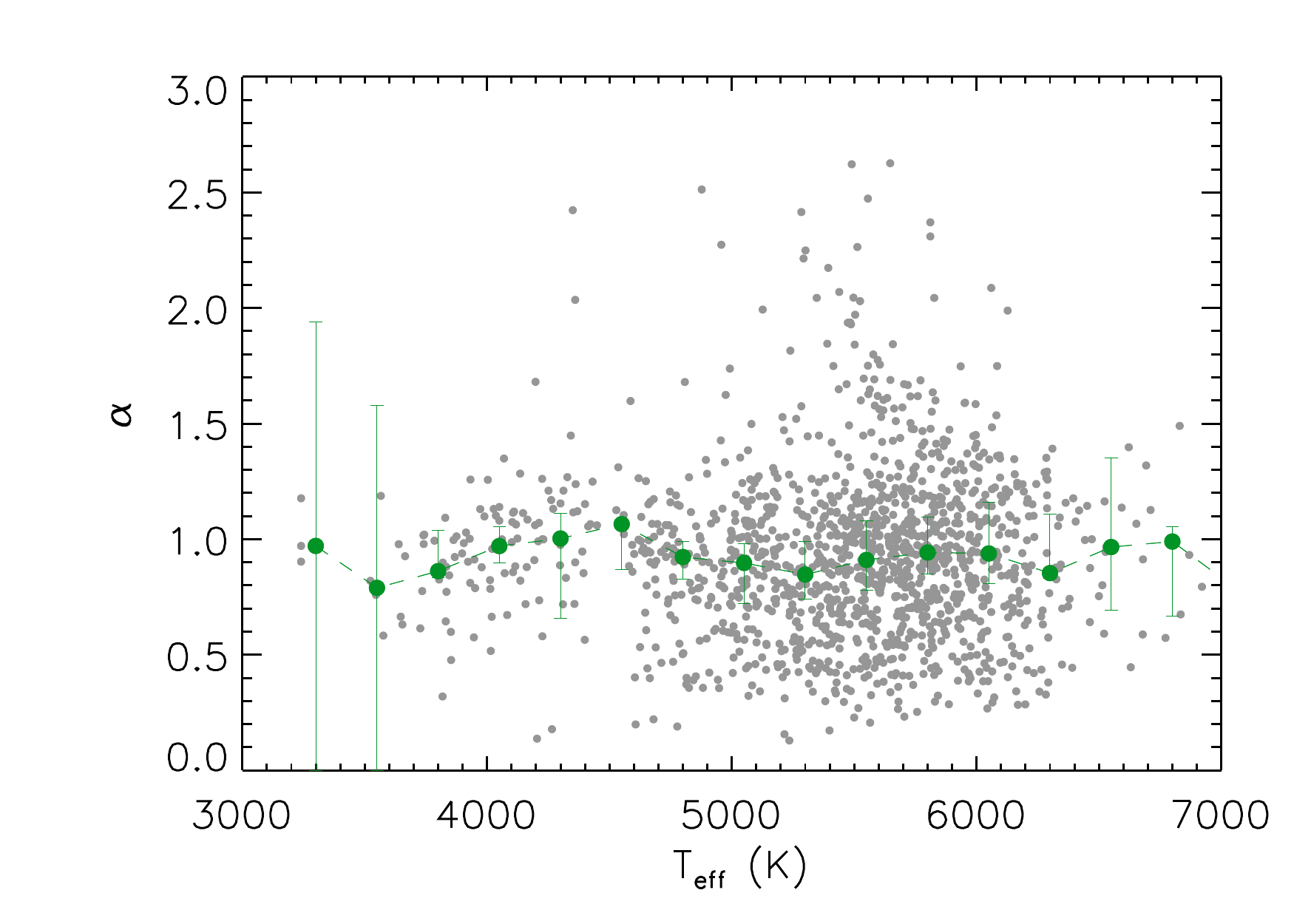}  
\includegraphics[scale=0.45,clip=true,trim=0cm 0cm 0cm 0cm]{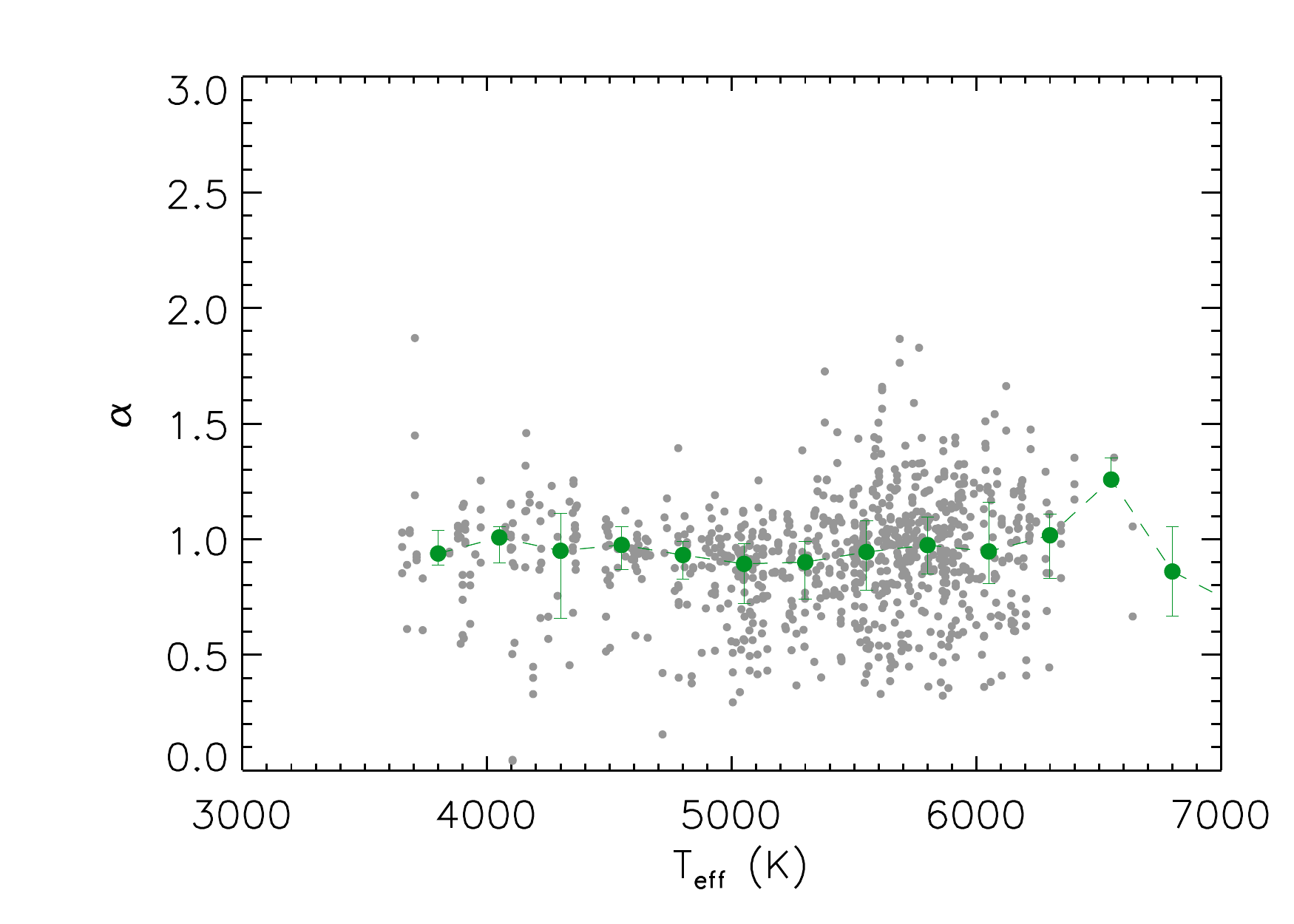}  
\caption{Same as Figure 4 for stellar effective temperature.}
\end{figure}

\begin{figure}
\centering
\includegraphics[scale=0.4]{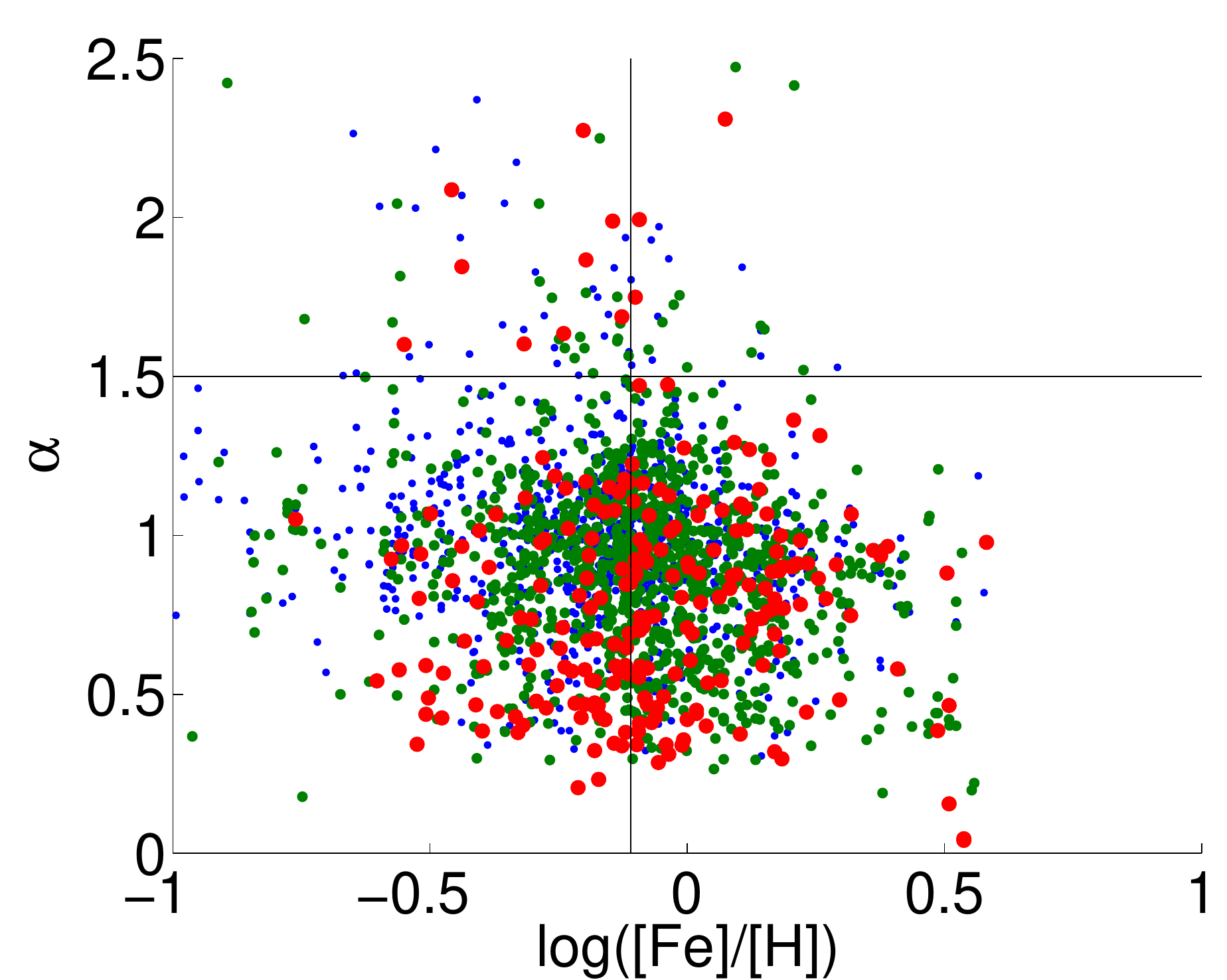}  
\includegraphics[scale=0.4]{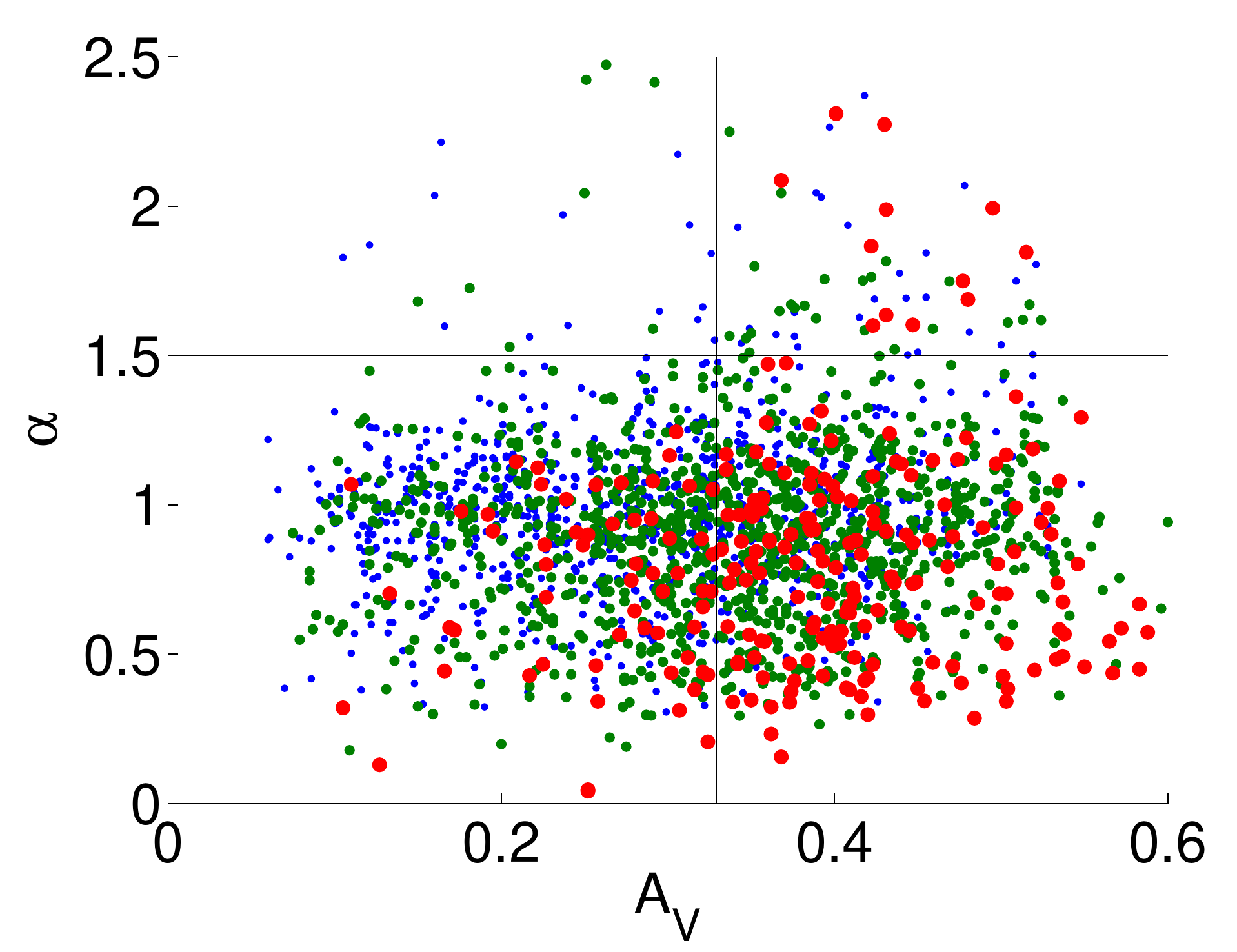}  
\caption{Distributions for the transit duration anomalies $\alpha$ as a function of stellar metallicity on the left and extinction on the right.  KOIs with exoplanet radii of $R_{pl}<2 R_{\oplus}$, $2 R_{\oplus}<R_{pl}<6 R_{\oplus}$, and $R_{pl}>6 R_{\oplus}$ are shown in blue, green and red respectively.  The horizontal line at $\alpha=1.5$ indicates that most  KOIs with $\alpha>1.5$ likely have large systematic errors in stellar parameters or are false positives, and are preferentially found with low metallicity and high extinction.  The vertical lines correspond to the median values for all KOIs in metallicity (-0.11) and extinction (0.33).}
\end{figure}

\begin{figure}
\centering
\includegraphics[scale=0.4]{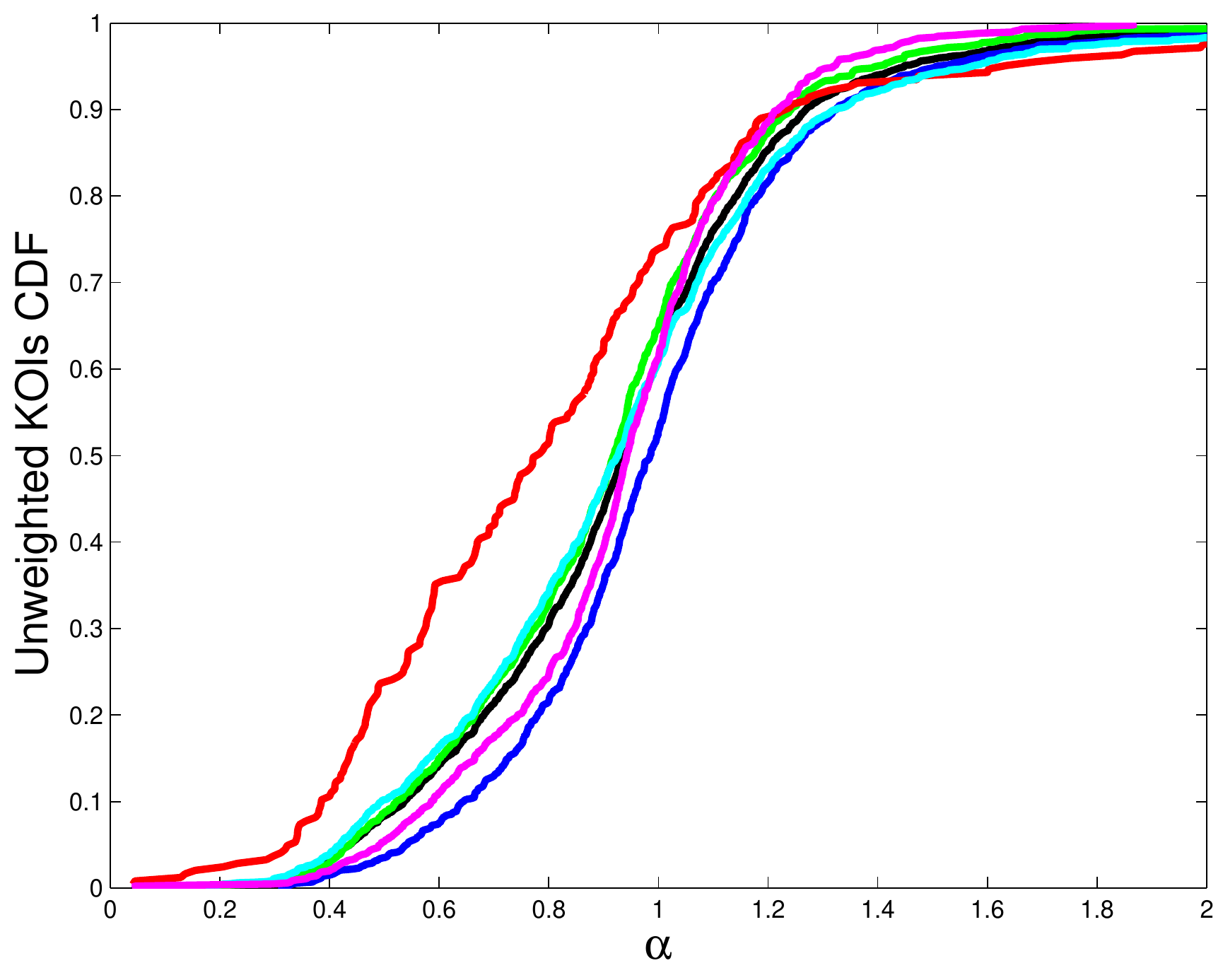}  
\includegraphics[scale=0.4]{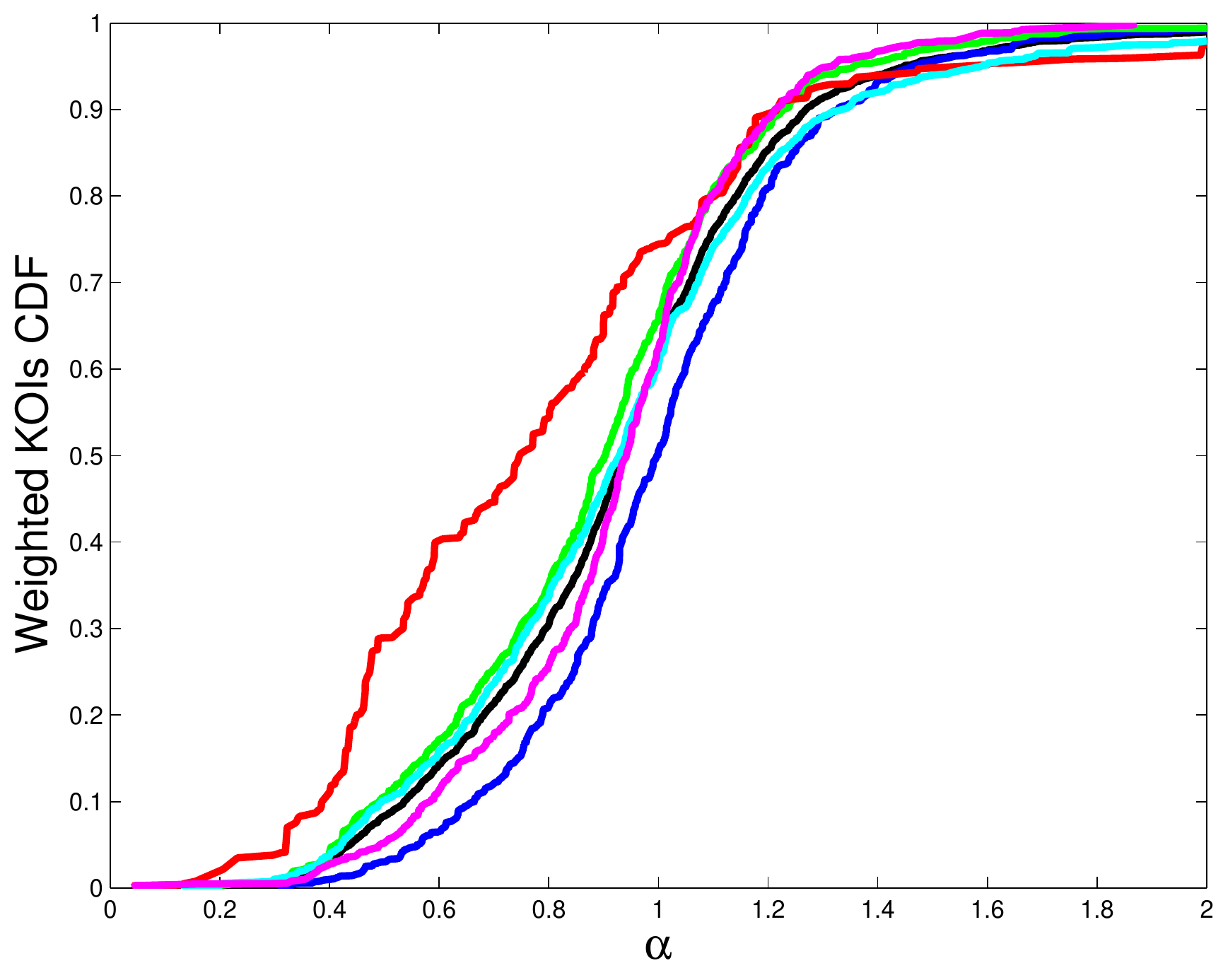}  
\caption{Left: CDF($\alpha$) curves of KOIs and subsets thereof, which are the same as the black curves shown in Figure 2:  Black: All KOIs from B13; Blue: $R_{pl}<2 R_{\oplus}$ KOIs;  Green: $2 R_{\oplus}<R_{pl}<6 R_{\oplus}$KOIs; red: $R_{pl}>6 R_{\oplus}$ KOIs; cyan: single KOIs; magenta: multi KOIs.  Right: The same as the left panel, except all the CDFs are corrected (weighted) to have matching period distributions -- matched to the period distribution of all KOIs -- to account for period-dependent eccentricity variations.  The same colors are used for the same KOI subsets in Figure 1.  See ${\S}$6.3 for discussion.}
\end{figure}

\end{document}